\documentclass[twocolumn]{aastex701}

\begin{document}

\title{Stratification of the AGN-Driven multi-phase outflows in the  dwarf Seyfert galaxy NGC 4395}

\correspondingauthor{Payel Nandi}

\author[orcid=0009-0003-9765-3517]{Payel Nandi}
\affiliation{Inter-University Centre for Astronomy and Astrophysics, IUCAA, Pune 411007, India}
\affiliation{Indian Institute of Astrophysics, Block II, Koramangala, Bangalore 560034, India}
\email[show]{nandipayel05@gmail.com}

\author[orcid=0000-0002-9090-4227]{Luis Colina}
\affiliation{Centro de Astrobiolog\'ia (CAB), CSIC-INTA, Ctra. de Ajalvir km 4, Torrej\'on de Ardoz, E-28850, Madrid, Spain}
\email{colina@cab.inta-csic.es}

\author[orcid=0000-0003-0483-3723]{Rogemar A. Riffel}
\affiliation{Departamento de F\'isica, CCNE, Universidade Federal de Santa Maria, 97105-900, Santa Maria, RS, Brazil}
\affiliation{Centro de Astrobiolog\'ia (CAB), CSIC-INTA, Ctra. de Ajalvir km 4, Torrej\'on de Ardoz, E-28850, Madrid, Spain}
\email{rogemar@ufsm.br}

\author[orcid=0000-0002-4005-9619]{Miguel Pereira Santaella}
\affiliation{Instituto de F\'isica Fundamental, CSIC, Calle Serrano 123, 28006 Madrid, Spain}
\email{miguel.pereira@iff.csic.es}

\author[orcid=0000-0002-4998-1861]{C.\ S.\ Stalin}
\affiliation{Indian Institute of Astrophysics, Block II, Koramangala, Bangalore 560034, India}
\email{stalin@iiap.res.in}

\author[orcid=0000-0002-4464-8023]{D.\ J.\ Saikia}
\affiliation{Fakult\"at f\"ur Physik, Universit\"at Bielefeld, Postfach 100131, D-33501 Bielefeld, Germany}
\affiliation{Assam Don Bosco University, Guwahati 781017, Assam, India}
\email{dhrubasaikia.tifr.ccsu@gmail.com}

\author[orcid=0000-0002-7093-1877]{Javier {\'A}lvarez-M{\'a}rquez}
\affiliation{Centro de Astrobiolog\'ia (CAB), CSIC-INTA, Ctra. de Ajalvir km 4, Torrej\'on de Ardoz, E-28850, Madrid, Spain}
\email{jalvarez@cab.inta-csic.es}

\author[orcid=0000-0002-5908-1488]{Markus Kissler-Patig}
\affiliation{European Space Agency (ESA), European Space Astronomy Centre (ESAC), Camino Bajo del Castillo s/n, 28692 Villanueva de la Ca{\~n}ada, Madrid, Spain}
\email{markus.kissler-patig@esa.int}

\begin{abstract}

We present a multi-wavelength study of nuclear outflows in the nearby dwarf Seyfert galaxy NGC~4395, which hosts an intermediate-mass black hole. Using \textit{JWST}/NIRSpec and MIRI IFU spectroscopy (1.66--28.6~$\mu$m), together with ALMA and Gemini/GMOS data, we probe the ionised and molecular gas on parsec scales. The JWST nuclear spectra reveal 134 emission lines, including H\,\textsc{i}, He, numerous fine-structure lines, H$_2$ rotational/ro-vibrational transitions, and several PAH bands.
Modelling of the H$_2$ rotational lines reveals three warm/hot molecular components ($T\!\approx\!580$, 1480, and 2900~K), along with a cold ($<50$~K) phase traced by ALMA CO(2--1). Outflow signatures are detected in cold and warm/hot molecular gas, in H\,\textsc{i}, and in 36 fine-structure lines spanning ionisation potentials of 7.6--300~eV. Ionised outflow velocities range from 127 to 716~km\,s$^{-1}$, with blueshifted and redshifted components consistent with a stratified biconical geometry.
The cold molecular gas shows a mass outflow rate nearly 1--2 orders of magnitude larger than that of the warm/hot molecular and ionised phases. The kinetic coupling efficiency is 0.003--0.12\% for the coronal-line gas and 0.4--1.4\% for the H\,\textsc{i} outflow, indicating that only the low-ionisation gas significantly impacts the surrounding ISM. Outflow velocity and the fraction of flux in the outflowing component increase with ionisation potential, implying that the most highly ionised gas originates closest to the AGN and is most efficiently accelerated.

\end{abstract}


\section{Introduction}

Active galactic nuclei (AGN) exert a significant influence on the formation and evolution of galaxies through powerful feedback mechanisms. These processes, driven either by radiation pressure or mechanical outflows such as radio jets, can regulate star formation, heat the interstellar medium (ISM), and redistribute gas across a wide range of spatial scales, from the inner few parsecs to kiloparsec sized galactic structures \citep{2012ARA&A..50..455F, 2015ARA&A..53..115K, 2018NatAs...2..198H, 2020A&ARv..28....2V, 2023ApJ...950...81N, 2023ApJ...959..116N, 2024ApJ...973....7N, 2024BSRSL..93..780N, 2024ApJ...974..195Z, 2024A&A...690A.350H}. 

AGN outflows are inherently multiphase, observed in ionised, neutral, and molecular gas, each tracing different regions and physical conditions of the ISM \citep{2023MNRAS.521.1832R, 2023MNRAS.520.5712S}. These outflows span a wide velocity and spatial range—from ultra-fast X-ray winds near the accretion disk \citep{2010A&A...521A..57T}, to extended molecular and ionised gas flows observed at kiloparsec scales \citep{2014MNRAS.441.3306H, 2023Sci...382..554I}.

Ultra-fast outflows (UFOs), often detected in X-ray spectra through blue-shifted absorption of Fe\,{\sc xxv} or Fe\,{\sc xxvi}, exhibit velocities $\gtrsim$10,000~km~s$^{-1}$ and are thought to originate within a few hundred gravitational radii from the central black hole \citep{2010A&A...521A..57T}. Intermediate UV absorbing outflows observed with HST show blueshifted resonance lines (e.g., C\,{\sc IV}, O\,{\sc VI}) with velocities of up to $\leq$3000–4000 km s$^{-1}$, tracing photoionized winds that bridge the gap between ultra‑fast X‑ray winds and the slower warm ionized gas phases in many Seyferts and quasars \citep{2004ApJ...612..152C, 2015A&A...577A..37A, 2021MNRAS.503.1442M}.
Warm ionized gas, traced through optical and near-infrared forbidden lines such as [O\,{\sc iii}]5007 or high-ionization coronal lines like [Fe\,{\sc vii}] and [Ne\,{\sc v}], exhibits typical velocities in the range of 500$-$2000~km~s$^{-1}$ and can extend to several kiloparsecs \citep{2002ApJ...579..214R, 2011ApJ...739...69M, 2025MNRAS.538.2800R}. 

In contrast, molecular outflows trace the cooler gas reservoirs and are observed via CO, OH, or H$_2$ emission or absorption, including both the cold ($T < 100$~K) and warm/hot ($T \sim 100-$3000~K) phases. While these outflows are typically slower (a few hundred to $\sim$1000~km~s$^{-1}$), they often carry significant mass and momentum and are directly linked to the suppression or triggering of star formation \citep{2014A&A...562A..21C, 2011A&A...533L..10D, 2014A&A...572A..40E, 2017A&A...607A.116E, 2017A&A...601A.143F, 2018A&A...616A.171P, 2020A&A...643A..89P, 2024ApJ...974..127C, 2024A&A...689A.314L}. Additionally, outflows in neutral atomic gas are seen in H\,{\sc i}\ 21~cm absorption or optical Na\,{\sc i}\,D lines, offering further insights into AGN-driven feedback processes on different scales \citep{2010ApJ...708.1145K, 2016A&A...590A.125C, 2018A&ARv..26....4M, 2021MNRAS.503.4748R, 2024MNRAS.528.4976D}.

High-ionization coronal lines, such as [Ne\,{\sc v}], [Fe\,{\sc vii}], and [Mg\,{\sc viii}], offer a unique probe of the innermost energetic regions of the AGN, believed to reside between the classical narrow-line region and the broad-line region \citep{1997ApJS..110..287F, 2023A&A...672A.108A}, where gas may be predominantly shock-heated or photoionized by the AGN continuum. However, such lines remain less commonly studied in outflow diagnostics compared to low-ionisation species \citep{2002ApJ...579..214R, 2011ApJ...739...69M, 2024A&A...685L..13P, 2023MNRAS.524..143F, 2025MNRAS.538.2800R}.

Disentangling the excitation mechanisms and kinematic signatures of these different gas phases is essential for understanding how AGN feedback regulates galaxy growth. Molecular gas, in particular, plays a critical role since it fuels star formation and can be strongly influenced by AGN activity. A holistic, multi-phase view is thus key to capturing the full extent and impact of AGN feedback \citep{2017A&A...601A.143F, 2021MNRAS.506.2950R}. While some studies have examined multi-phase outflows, most of them have focused on massive galaxies hosting luminous quasars \citep{2013ApJ...768...75R} or Seyfert nuclei such as in Cygnus A, NGC~424 \citep{2021MNRAS.506.2950R, 2025arXiv250321921M}, there is increasing interest in exploring feedback in low-mass galaxies hosting intermediate-mass black holes (IMBHs, $M_{\rm BH} \sim 10^4-10^6\,M_\odot$). These systems are crucial for understanding early black hole seed formation, low-mass end coevolution, and the effectiveness of AGN feedback in shallow gravitational potentials \citep{2015ApJ...813...82R, 2018MNRAS.476..979P, 2020ApJ...899L...9G}. However, observational studies of such systems remain sparse, primarily due to their lower luminosities and compact sizes, which challenge both spatial and spectral resolution.

NGC~4395, located at 4.3 Mpc \citep{2004AJ....127.2322T}, is one of the best studied dwarf AGN with an IMBH of mass $\sim 10^5$~M$_\odot$ \citep{2024ApJ...976..116P, 2015ApJ...809..101D, 2005ApJ...632..799P, 2003ApJ...588L..13F}. Despite its low bolometric luminosity ($L_{\rm bol} \sim 10^{41}$ erg s$^{-1}$), NGC~4395 shows pronounced nuclear activity, hosting hot ionized outflows traced by X-ray and UV absorbers \citep{2004ApJ...612..152C, 2012ApJ...753...75C}, as well as warm ionized outflows traced by optical [O\,\textsc{III}], which may be driven by a compact radio jet \citep{2023ApJ...959..116N}.
 Additionally, evidence for both negative and positive feedback modes has been reported on sub-kpc scales \citep{2023ApJ...950...81N, 2023ApJ...959..116N}.

In this work, we analyzed JWST/NIRSpec and MIRI IFU spectra, complemented by ALMA and Gemini/GMOS observations, to carry out a comprehensive study of nuclear outflows in NGC~4395 across the near- and mid-infrared, molecular, and optical bands. This paper is structured as follows: Section~\ref{sec:data-ana} describes the observations and data reduction; Section~\ref{sec:result} presents the main results, including emission-line diagnostics and kinematic profiles; Section~\ref{sec:discussion} discusses the multiphase nature of the outflows, their dependence on ionization potential (IP: the energy required to produce the ionization stage responsible for the observed line), and the implications for AGN feedback in dwarf galaxies; and Section~\ref{sec:summary} summarizes the key conclusions. Throughout this paper we adopt a distance of 4.3~Mpc to NGC~4395 \citep{2004AJ....127.2322T}, corresponding to a scale of 21 pc arcsec$^{-1}$.  


\section{observations and data reduction and analysis} \label{sec:data-ana}
\subsection{JWST}
We utilised archival \textit{integral field unit} (IFU) observations from the \textit{JWST} obtained under \text{Program ID 2016 (PI: Anil C. Seth) \citep{2026arXiv260116977G}}. These observations were conducted using JWST's two primary spectroscopic instruments: the \textit{Near Infrared Spectrograph} (NIRSpec; \citealt{2022A&A...661A..80J}) and the \textit{Mid-Infrared Instrument} (MIRI; \citealt{2015PASP..127..595W, 2015PASP..127..584R, 2023PASP..135d8003W}). 

For the \textit{NIRSpec IFU} observations, we used data acquired in high-resolution mode with the \texttt{G235H} and \texttt{G395H} gratings, coupled with the \texttt{F170LP} and \texttt{F290LP} filter configurations, respectively. These configurations collectively cover the 1.66--5.2~$\mu$m wavelength range at a spectral resolving power $R \approx 1000-5000$ \citep{2023PASP..135c8001B}. These observations were performed using a 4-point dither strategy, enhancing spatial coverage and minimising residual detector-related artifacts such as bad pixels and inter-pixel capacitance.

For MIRI, the IFU data span all four spectral channels (Ch1--Ch4), each subdivided into three spectral sub-bands: short, medium, and long, resulting in a total of 12 spectral segments. These observations were also performed using a 4-point dither strategy. MIRI’s IFU mode provides moderate spectral resolution ($R \approx$ 1500$-$3500) across the 4.9--28.6~$\mu$m wavelength range \citep{2021A&A...656A..57L, 2023A&A...675A.111A, 2023MNRAS.523.2519J}.

The raw uncalibrated data were downloaded from the Barbara A. Mikulski Archive for Space Telescopes (MAST) portal and then reduced using the official \texttt{jwst} calibration pipeline (version 1.17.2), along with CRDS reference files from CRDS context \texttt{1322.pmap}, \citep{2016A&C....16...41G}. For both instruments, we employed \texttt{Stage~1} and \texttt{Stage~2} of the pipeline, which perform detector-level corrections and spectrophotometric calibrations.

For MIRI, we implemented an additional step to identify and mask warm pixels, using background exposures from \texttt{Stage~2} products to generate pixel masks. These were applied to the corresponding science frames before proceeding to \texttt{Stage~3}, where the final three-dimensional data cubes were constructed using \texttt{Drizzle} algorithm \citep{2023AJ....166...45L}. During \texttt{Stage~3}, we enabled master background subtraction to remove residual thermal and instrumental background contributions, benefiting from dedicated background exposures available for these observations.

In the case of NIRSpec, we similarly employed \texttt{Stage~1} and \texttt{Stage~2} of the pipeline. Additionally, to address the presence of 1/$f$ noise, we applied the \texttt{NSClean} algorithm during \texttt{Stage~2}. As the NIRSpec program did not include separate background observations, no background subtraction was performed for these data. Then, we ran \texttt{Stage~3} of the pipeline to combine all the exposures and construct the cube using the \texttt{Drizzle} method. We also rescaled our pixel to 0$\farcs$05.

From the fully calibrated IFU data cubes, we extracted the nuclear spectra of the central AGN using a circular aperture with a radius of 0.5$^{\prime\prime}$ for the NIRSpec gratings and MIRI Channels~1 and 2. Owing to the increasing degradation of the point spread function (PSF) at longer wavelengths, a larger aperture of 1$^{\prime\prime}$ was adopted for MIRI Channels~3 and 4 to account for the extended PSF wings. To ensure accurate flux calibration within each aperture, we performed wavelength-dependent PSF simulations using the \texttt{Space Telescope PSF (STPSF)} package \citep{2012SPIE.8442E..3DP, 2014SPIE.9143E..3XP}, which incorporates instrumental effects such as geometric distortion and detector charge-transfer inefficiencies. The extracted spectra were corrected for aperture flux losses using the corresponding PSF models, enabling reliable recovery of the intrinsic AGN flux across the full near- to mid-infrared spectral range.  

A relative offset of $2\times10^{-18}$~erg~s$^{-1}$~cm$^{-2}$ was identified between all channels and Channel~4. In addition, Channel~4C exhibited an extra deviation of $5\times10^{-18}$~erg~s$^{-1}$~cm$^{-2}$ relative to Channels~4A and 4B. These discrepancies were corrected by applying vertical shifts: Channels~4A and 4B were shifted downward by $2\times10^{-18}$~erg~s$^{-1}$~cm$^{-2}$, while Channel~4C was shifted upward by $5\times10^{-18}$~erg~s$^{-1}$~cm$^{-2}$.

\subsection{ALMA}
To probe the cold molecular gas component of the outflows, we utilised high spectral resolution observations from the \textit{Atacama Large Millimeter/submillimeter Array} (ALMA). Specifically, we analysed archival \texttt{Band~6} observations obtained with the 12-m array, under project ID \texttt{2017.1.00572.S} (PI: Timothy Davis). We utilized the pipeline-calibrated and primary beam-corrected data cube available from the ALMA Science Archive, which had been processed using \texttt{CASA} version \texttt{5.4.0-68}. The full-width half maximum (FWHM) of the synthesised beam is $1.99^{\prime\prime} \times 1.29^{\prime\prime}$. The spectral resolution is uniformly high across the band, with channel spacing of 0.63--0.64~km~s$^{-1}$, enabling precise characterisation of narrow molecular line components. Spectral extraction was performed using an aperture matched to the beam size, centred on the flux-weighted centroid of the emission as derived from the velocity-collapsed line map which is $\sim0.9^{\prime\prime}$ far from the centre of the nucleus towards northward west direction as reported in \cite{2023ApJ...959..116N}.

\subsection{GMOS}
In addition to millimeter and infrared observations, we incorporated optical IFU spectroscopy obtained with the \textit{Gemini Multi-Object Spectrograph} (GMOS) on the \textit{Gemini-North} telescope  to use optical coronal lines [Fe\, {\sc VII}] for our density-temperature analysis. These data were obtained as part of program \text{GN-2015A-DD-6} (PI: Rachel Mason), and cover the central region of the galaxy with spectral range of 4450--7347~\AA. From the final reduced GMOS data cube (see \citealt{2019MNRAS.486..691B} and \citealt{2023ApJ...959..116N} for a detailed description of the observations and data reduction procedures), we extracted the integrated optical spectrum within a circular aperture of radius 0$\farcs$5 centred on the nucleus, matching the apertures used for the infrared observations.

\begin{figure*}
    \centering
        \includegraphics[scale=0.55]{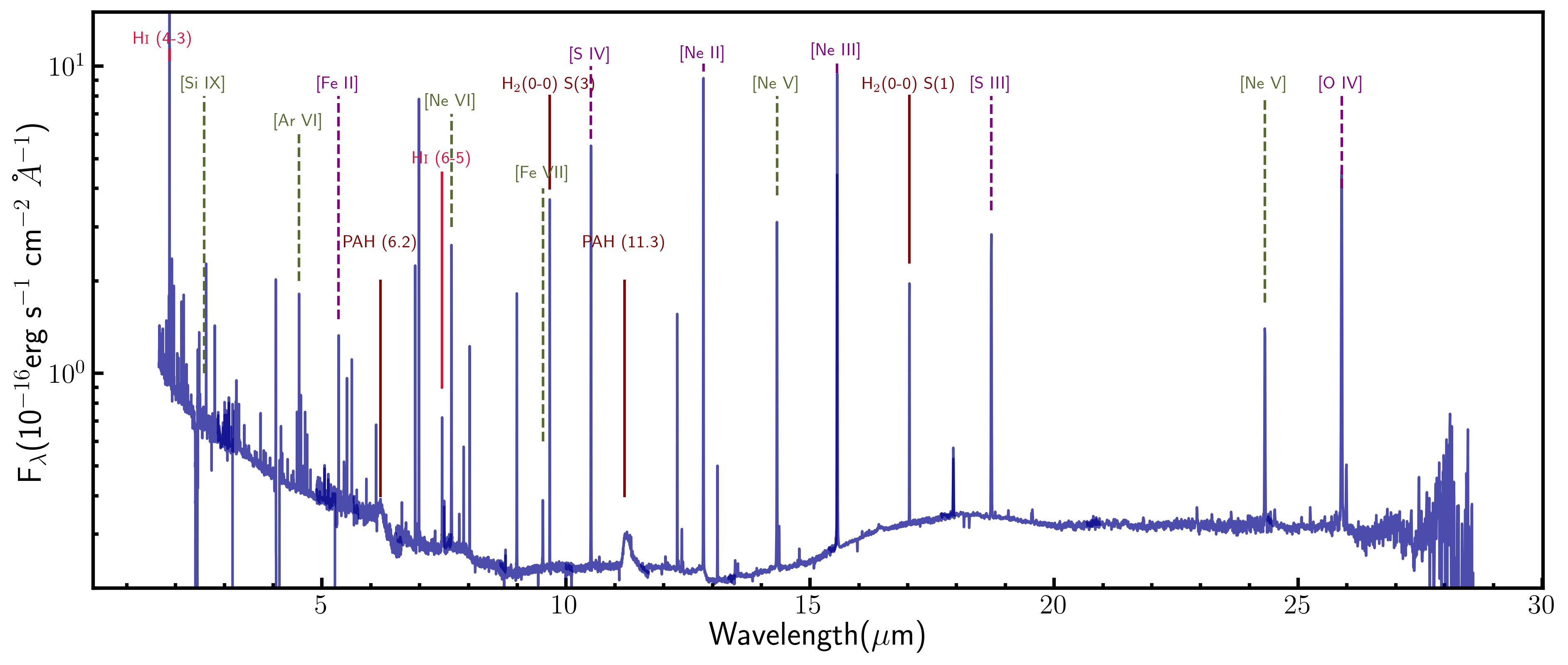}
        \caption{The aperture-corrected nuclear spectrum of NGC\,4395, extracted from the central region, spans the rest-frame wavelength range 1.66--28.6~$\mu$m, combining observations from \textit{JWST}/NIRSpec and MIRI Channels~1--4. Some of prominent emission lines are indicated along the spectrum. Molecular features, including PAH and H$_2$ lines, together with H$\textsc{i}$ recombination lines, are shown by solid lines, while ionized gas lines are denoted by dashed lines. Low-ionization species (e.g., [Fe\,\textsc{ii}], [Ne\,\textsc{ii}], [S\,\textsc{iii}], [O\,\textsc{iv}]) are marked in purple, whereas high-ionization lines are indicated in dark green.
}

    \label{fig:spectra}
\end{figure*}

\section{Results}\label{sec:result}

\subsection{Identification of lines and their characteristics} \label{sec:line-inst}
We extracted nuclear spectra from the \textit{JWST}/NIRSpec and MIRI instruments, covering the full spectral range from 1.66 to 28.6~$\mu$m. 

To systematically identify and characterise the spectral features in the extracted spectra, we utilised the Atomic Line List\footnote{\url{https://linelist.pa.uky.edu/atomic/}} and cross-referenced them with values reported in the literature. In total, we identified 134 distinct emission lines and polycyclic aromatic hydrocarbon (PAH) features across the combined NIRSpec and MIRI wavelength range. Of these, 91 were detected in the NIRSpec spectra and 50 in the MIRI spectra, with 7 lines appearing twice due to overlapping wavelength coverage between the NIRSpec G235H, NIRSpec G395H gratings and the MIRI sub-bands. The nuclear spectrum, with several of the prominent emission lines marked, is presented in Figure~\ref{fig:spectra}.

Following the identification of individual emission lines, we employed a multi-Gaussian fitting procedure to derive integrated fluxes and kinematic properties for each spectral feature. We first attempted a single-Gaussian fit for the line, combined with a first-order polynomial for the local continuum, and inspected the residuals. If the residuals within the line region exceeded those in the continuum by more than a factor of two, we added a second Gaussian component while retaining the polynomial continuum (see \citealt{2025ApJ...984...20N} for details). In practice, we found that a maximum of two Gaussians was sufficient to model all lines except three strong H\,\textsc{i} recombination lines. The full list of detected features, along with their measured fluxes and wavelengths, is provided in Table~\ref{tab:line-list}. This approach ensured a consistent treatment of blended or asymmetric profiles and enabled us to separate narrow and broad components in cases where the line shapes indicated the presence of outflowing gas.

During the line-fitting analysis, we noted that several emission lines in the MIRI spectra, especially those arising in Channels~3 and 4, exhibited FWHM values that were narrower than the nominal instrumental resolution as determined from in-flight performance measurements \citep{2023A&A...675A.111A}. To assess the reliability and physical plausibility of these narrow features, we cross-validated our FWHM measurements with recent studies of high-resolution MIRI spectra in circumstellar disk systems \citep{2024ApJ...963..158P, 2025AJ....169..165B}. These works reported similarly narrow emission lines, suggesting that, when instrumental resolution is properly accounted for and a conservative uncertainty budget (10\%) is included, such measurements remain credible. Our comparison, illustrated in Figure~\ref{fig:res}, shows good agreement with these prior results and supports the interpretation that narrow, unresolved or marginally resolved lines can persist in AGN environments, particularly in low-luminosity systems where turbulent and bulk velocities may be modest.

Furthermore, we observed that the emission lines detected in the NIRSpec spectra also displayed FWHM values that were systematically smaller than those expected from ground-based calibration observations. However, when compared against reported uncertainties in ground-based instrumental resolution, our measurements remained within the 30\% error range cited in the literature. This consistency, demonstrated in Figure~\ref{fig:res}, provides additional confidence in the robustness of our line-fitting methodology and supports the physical interpretation of the observed line widths.


Importantly, for all lines in both MIRI and NIRSpec, in which an outflow component was identified, the FWHM of that component exceeded the instrumental resolution, and we corrected for instrumental broadening in all outflow-related measurements.

\subsubsection{Hydrogen Recombination Lines} \label{sec:HI-lines}
    We identified multiple hydrogen recombination (H\,\textsc{i}) lines (a total of 24 lines) across the Paschen (Pa), Brackett (Br), Pfund (Pf), and Humphreys (Hu) series in the near- and mid-infrared spectral range, which trace photoionized gas associated with both AGN activity and circumnuclear star formation. Among the Paschen lines, only the Pa$\alpha$ transition was detected within the wavelength coverage of our observations; it is the strongest H\,\textsc{i} line. 
    
    In the Br series, we detected six transitions, ranging from Br$\beta$ to Br$\eta$. The Br$\alpha$ line, which falls in the gap between detectors in the NIRSpec/G235H grating, was not covered and thus could not be measured. The Pf series is well-sampled, with eight lines detected from Pf$\alpha$ up to Pf(13$-$5), while in the Hu series we identified seven lines, extending from Hu$\alpha$ to Hu$\eta$.

     In addition to these classical series, we also detected two higher-order hydrogen recombination transitions. These include one line from levels with lower principal quantum number $n=7$, specifically H\,\textsc{i} (11–7) with a signal-to-noise ratio (SNR) of 3.77. We also report the detection of a line at 12.387~$\mu$m corresponding to a transition between the $n=11$ and $n=8$ energy levels, with a SNR of 3.07. 
     
     
    Most of the H\,\textsc{i} lines (except for weak transitions such as Pf(13$-$5), Hu(13$-$6), H\,\textsc{i}(11$-$7), and H\,\textsc{i}(11$-$8)) are best fitted with two Gaussian components, including a blueshifted broad component. 
    For the stronger H\,\textsc{i} lines, such as Pa$\alpha$, Br$\beta$, and Br$\gamma$, where the SNR exceeds 40, we were able to clearly disentangle the contribution from the very broad (FWHM$>$1300 km s$^{-1}$) Broad Line Region (BLR) component. These lines were modelled with three Gaussian components: one representing the Narrow Line Region (NLR), another for the BLR (sharing the same systemic velocity as the NLR), and a third blueshifted component accounting for the outflow emission, following the approach of \cite{2023ApJ...959..116N}. We then only considered NLR and outflow component for our further analysis.
    
\subsubsection{Helium Lines} 
We identified eight helium emission lines in the NIRSpec spectrum, including six He\,\textsc{ii} recombination lines and two neutral He\,\textsc{i} recombination transitions. Most of these lines were well modelled with a single Gaussian component. Only He\,\textsc{i}\,1.869\,$\mu$m and 
He\,\textsc{ii}\,4.7635\,$\mu$m required two components; however, the broad-line region (BLR) component could not be reliably resolved due to the modest SNR (S/N~$<40$) of these features. These helium lines originate in highly ionised gas and provide important diagnostics for probing 
the hardness of the ionising radiation field.

\subsubsection{Fine-Structure Atomic Lines} 
    We detected a rich set of forbidden emission lines (total 61 lines) from a wide array of ionized species, including heavy elements such as [O\,\textsc{iv}], [Ne\,\textsc{ii}], [Ne\,\textsc{iii}], [Ne\,\textsc{v}], [Ne\,\textsc{vi}], [Ni\,\textsc{ii}], [Na \,\textsc{iii}], [Na \,\textsc{vii}], [Mg\,\textsc{v}], [Mg\,\textsc{vii}], [Mg\,\textsc{viii}], [Si\,\textsc{ix}], [S\,\textsc{iii}], [S\,\textsc{iv}], [Ar\,\textsc{ii}], [Ar\,\textsc{iii}], [Ar\,\textsc{v}], [Ar\,\textsc{vi}], [Ca\,\textsc{iv}], [Cl\,\textsc{v}], [K\,\textsc{iii}], [Fe\,\textsc{ii}], [Fe\,\textsc{iii}], [Fe\,\textsc{vii}], [Co\,\textsc{ii}]. These lines span a broad range of IP, from $\sim$7.6~eV (e.g., [Ni\,\textsc{ii}]) to $\sim$303~eV (e.g., [Si\,\textsc{ix}]), enabling a comprehensive investigation of the ionised gas conditions in the nuclear region.
    
    
    Low- and intermediate-ionization lines (e.g., [S\,\textsc{iii}], [Ne\,\textsc{ii}], [Fe\,\textsc{ii}]) trace photoionized regions associated with the extended NLR, while high-ionization lines (e.g., [Ne\,\textsc{v}], [Fe\,\textsc{vii}], [Si\,\textsc{ix}]) are unambiguous tracers of AGN activity and can also be excited by fast ($v \gtrsim 200$ km s$^{-1}$) radiative shocks \citep[e.g.,][]{2008ApJS..178...20A, 2017ApJS..229...34S}.
    

\subsubsection{Molecular Hydrogen (H$_2$) Lines} 
A comprehensive set of 41 molecular hydrogen (H$_2$) emission lines, including both rotational and ro-vibrational transitions, was detected using the NIRSpec, and MIRI data between 1.75 to 17.04 $\mu$m. These lines provide valuable insights into the properties of warm and hot molecular gas in the nuclear region.

We identified 15 pure rotational lines from the 0--0 S-branch, ranging from H$_2$ S(1) at 17.035~$\mu$m to H$_2$ S(15) at 3.6263~$\mu$m. Two lines from the 1--1 S-branch, H$_2$ S(9) and H$_2$ S(10), were also detected; although, H$_2$(1--1)~S(10) is blended with the strong H\,\textsc{i} recombination line Pf$\beta$, making its detection uncertain.

From the v=1--0 ro-vibrational band, we detected sixteen lines: eight in the S-branch (H$_2$~S(0) to S(7)), five in the O-branch, and three in the Q-branch (H$_2$~Q(5), Q(6), and Q(7)). The Q(1) to Q(4) lines fall in a detector gap of the NIRSpec/G235H grating and could not be observed. The five detected O-branch transitions range from H$_2$(1--0)~O(3) at 2.8025~$\mu$m to H$_2$(1--0)~O(7) at 3.8074~$\mu$m. H$_2$(1--0)~O(2), which lies at 2.6269~$\mu$m, is blended with Br$\beta$, complicating its detection.

We also detected higher-excitation ro-vibrational lines from the v=2--1 band, including two S-branch lines (H$_2$~S(0) and S(1)) and four O-branch lines (H$_2$~O(3) to O(6)). Additionally, two O-branch transitions from the v=3--2 band, H$_2$~O(3) and O(5), were identified.

Some of H$_2$ emission lines are narrow, with widths comparable to the instrumental resolution, indicating that the emitting gas is kinematically quiescent. The excitation of this warm and hot molecular gas traced through these pure rotational and ro-vibrational transitions, with temperatures ranging from a few hundred to a few thousand K, is likely driven by a combination of ultraviolet fluorescence, X-ray heating, and shocks mechanisms commonly associated with AGN activity \citep{1994ApJ...427..777M}.    
    

\subsubsection{PAH features}
In addition to the emission lines shown in Fig.~\ref{fig:spectra}, we clearly detect prominent PAH features at 3.3~$\mu$m, 6.2~$\mu$m, and 11.3~$\mu$m. We also identify a broad hump between 15 and 20~$\mu$m, corresponding to the PAH plateau in this wavelength range \citep{2006ApJ...653..267T}. This feature likely arises from the blending of several weak PAH bands (e.g., 15.8, 16.4, 17.4, and 17.8~$\mu$m) emitted by large neutral or ionized PAH molecules, producing a smooth continuum-like ``hump'' \citep[e.g.,][]{2006ApJ...653..267T, 2007ApJ...656..770S, 2010A&A...511A..32B}.  

A possible additional contribution may come from thermal emission by warm dust grains in the torus or the circumnuclear ISM. Graphite and silicate grains with temperatures of 150--300~K can emit strongly in this spectral region, producing a broad mid-infrared continuum bump \citep[e.g.,][]{2004MNRAS.348.1065C, 2009ApJ...705..298M, 2015ApJ...803..110H}. In such cases, the 10~$\mu$m silicate feature often appears in emission (as seen in several type~1, unobscured AGN), with its long-wavelength tail extending toward $\sim$20~$\mu$m \citep{2017MNRAS.469..110G}. However, in our spectrum of NGC~4395, we do not detect silicate emission at 10~$\mu$m, while a strong 11.3~$\mu$m PAH feature is clearly present. This suggests that the observed 15--20~$\mu$m bump predominantly originates from the PAH plateau rather than from thermal dust emission.  

These broad PAH bands are attributed to the vibrational modes of large carbonaceous molecules that can survive in the harsh radiation field of the AGN, indicating the presence of resilient PAH populations in the circumnuclear environment \citep[e.g.,][]{10.1093/mnras/stz1316, 10.1093/mnras/stae1535, 2024A&A...691A.162G}.

\subsection{Emission-Line Widths and Kinematics}

\begin{figure}
    \centering
    \includegraphics[scale=0.35]{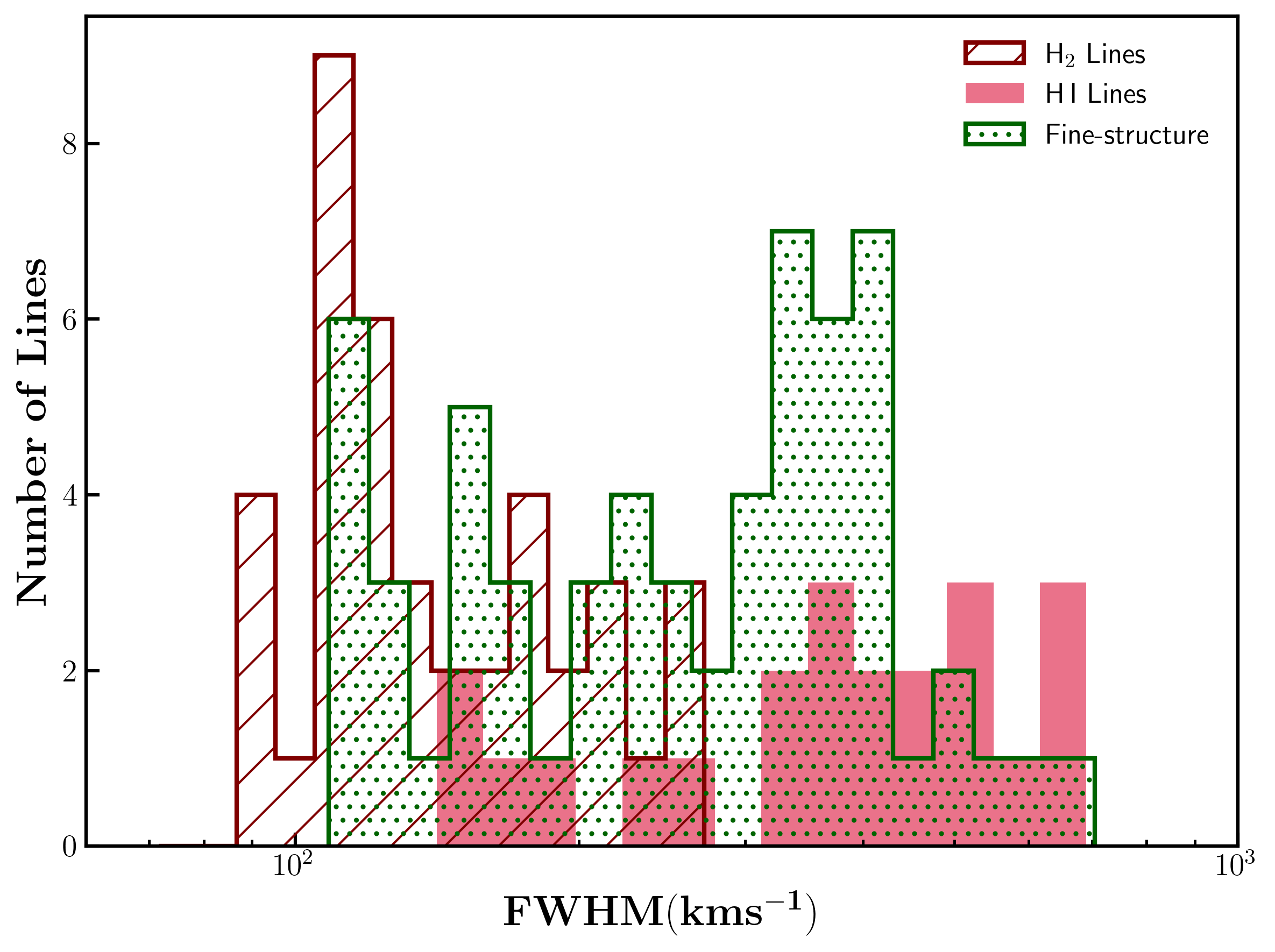}
    \caption{Distribution of the observed FWHM values (uncorrected for instrumental resolution) for the different gas phases: H$_2$, H \textsc{i}, and fine-structure lines.}
    \label{fig:dispersion}
\end{figure}

The distribution of the observed emission-line widths (FWHM) is presented in Fig.~\ref{fig:dispersion} for the molecular H$_2$, H\,\textsc{i}, and fine-structure lines.  
For the H\,\textsc{i} lines, we adopted the combined FWHM of the narrow and broad components for all transitions, and the sum of the narrow and broad outflow components for the three lines where a distinct BLR contribution could be fitted, as discussed in Section~\ref{sec:HI-lines}.  
The H\,\textsc{i} lines exhibit significant broadening, with FWHM values ranging from 141 to 690~km~s$^{-1}$ and a median of 422~km~s$^{-1}$.  
The wide spread in widths likely arises because some higher-order transitions appear narrower, possibly due to the absence of a detectable broad component from outflow and also BLR, which may in turn result from lower SNR.

In contrast, fine-structure lines that follow highly ionised species also exhibit a wide range of kinematic widths, with FWHM values ranging from 108 to 705~km~s$^{-1}$ and a median of 286~km~s$^{-1}$. 
These kinematic features are consistent with earlier studies of fine-structure lines in AGN, which show line broadening as a signature of outflowing ionised gas and stratified emission regions \citep{2025MNRAS.536.1166E, 2025ApJ...984...20N}.

Molecular hydrogen (H$_2$) emission lines exhibit the narrowest kinematic signatures among all detected species. The measured FWHM values span from 72 to 272 km s$^{-1}$, with a median of 126 km s$^{-1}$. However, three ro-vibrational H$_2$ (1--0) transitions—namely S(5), S(6), and S(7)—display tentative broad components with FWHM exceeding 400 km s$^{-1}$. Given the low signal-to-noise ratios and large uncertainties associated with these features, we exclude them from our statistical analysis to avoid biasing the FWHM distribution. The narrow widths of the lower-level H$_2$ lines are consistent with previous results from Gemini/NIFS observations of this system \citep{2019MNRAS.486..691B}. This range of line widths likely reflects combination excitation through UV fluorescence or thermal processes within rotating molecular structures, such as disks or compact tori or shocks \citep{1989ApJ...347..863S, 2005ApJ...633..105D}.

\subsection{Electron Density and Temperature Diagnostics}\label{sec:ne-Te}

We determined the electron density ($n_e$) and electron temperature ($T_e$) in the nuclear region of NGC~4395 through a detailed analysis of multiple low- and high-ionisation emission lines, utilising the \texttt{PyNeb} package \citep{2015A&A...573A..42L}. A variety of diagnostic line ratios were selected to sample a wide range of physical conditions that are insensitive to either $T_e$ or $n_e$, allowing for reliable and independent constraints on both parameters. We corrected all lines for extinction effect using \citet{2023ApJ...950...86G} reddening law and adopting $A_V = 0.96$ \citep{2019MNRAS.486..691B, 2023ApJ...959..116N}.

\begin{figure*}
    \centering
    \hbox{
    \includegraphics[scale=0.31]{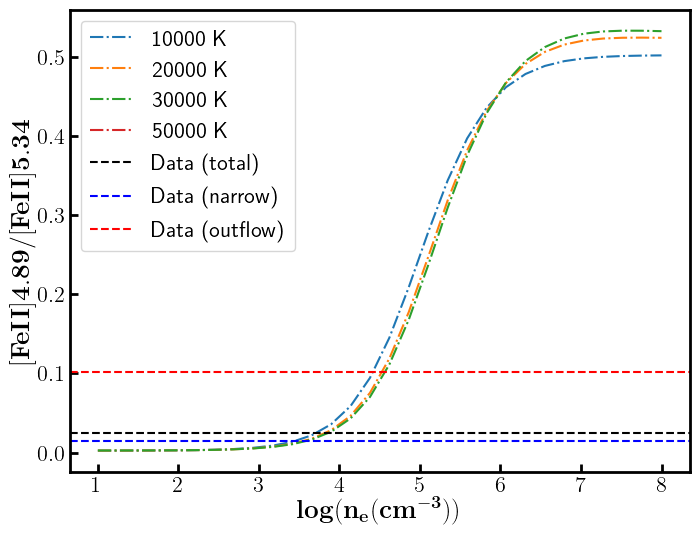}
    \includegraphics[scale=0.31]{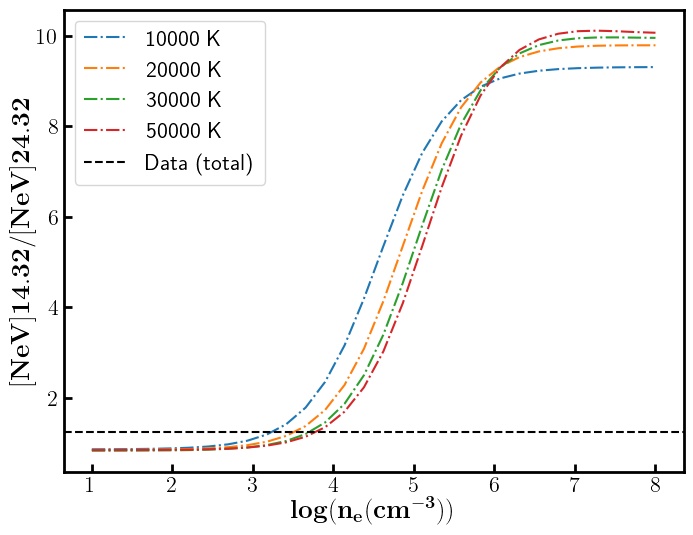}
    \includegraphics[scale=0.31]{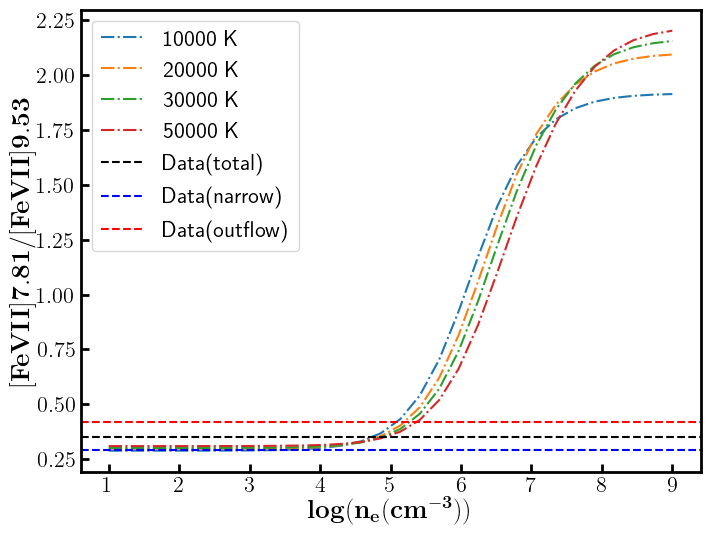}
    }
    \caption{Variation of different line ratios with density for different particular temperatures, which represent different curves in each plot and are denoted by individual legends in each plot. The horizontal lines in each plot represents the observed line ratio.}
    \label{fig:density}
\end{figure*}

To characterise the low-ionisation phase of the ionised gas, we analysed several prominent [Fe\,\textsc{ii}] emission line ratios. Among these, the [Fe\,\textsc{ii}]4.89/[Fe\,\textsc{ii}]5.34 ratio is particularly effective, as it shows strong sensitivity to electron densities in the range $10^{3}$–$10^{6}$~cm$^{-3}$, while exhibiting only a weak dependence on T$_e$ (see Fig.~\ref{fig:density}), consistent with the findings of \citet{2022A&A...665L..11P}. We evaluated this diagnostic across representative temperatures of 10,000\,K, 20,000\,K, 30,000K and 50,000\,K. From the observed [Fe\,\textsc{ii}] 4.89/5.34 ratio, we derived electron densities in the range $(0.50$–$0.82) \times 10^4$~cm$^{-3}$ ($n_e$ for individual components are given in Table \ref{tab:density}).

These values are significantly higher than those inferred from optical [S\,\textsc{ii}] doublet diagnostics in \citet{2023ApJ...959..116N}, which probe more extended, lower-density gas. Similar trends have been observed in other AGN \citep[e.g.][]{2021MNRAS.506.2950R}, highlighting that [Fe\,\textsc{ii}] emission lines are more responsive to shock-excited gas and originate from regions closer to the central engine, where densities are typically higher.

When separating the line emission into narrow and broad kinematic components, we found that the narrow components yield electron densities of $(0.27$–$0.43) \times 10^4$~cm$^{-3}$, in better agreement with values derived from [S\,\textsc{ii}] diagnostics \citep{2023ApJ...959..116N}. In contrast, the broad components exhibit considerably higher densities, ranging from $(2.58$–$4.25) \times 10^4$~cm$^{-3}$, indicating that the high-velocity outflowing gas is associated with denser regions likely located closer to the nucleus.

To probe the highly ionised gas component, we employed diagnostic transitions of [Ne\,\textsc{v}] and [Fe\,\textsc{vii}]. For [Ne\,\textsc{v}], we detected emission lines at 14.32\,$\mu$m and 24.32\,$\mu$m, enabling the use of the [Ne\,\textsc{v}]14.32/[Ne\,\textsc{v}]24.32 line ratio—a commonly adopted tracer of electron density in high-ionisation regions \citep{2008ApJ...676..836T, 2010ApJ...725.2270P, 2024ApJ...977..156D}. Based on the observed fluxes, we derived electron densities in the range of $(0.17$–$0.61) \times 10^4$~cm$^{-3}$ over the adopted temperature grid. In particular, due to the reduced sensitivity of the line and the limited spectral resolution and the relatively low intensity of the [Ne\,\textsc{v}]~24.32\,$\mu$m line, we were unable to resolve the outflow component. As a result, $n_e$ estimates for individual kinematic components could not be determined in this case.

We also investigated the physical conditions of the highly ionised gas phase using [Fe\,\textsc{vii}] transitions, with three prominent lines detected, two in the mid-infrared at 7.81\,$\mu$m and 9.53\,$\mu$m, and one in the optical at 6087\,\AA. Among the available diagnostics, the [Fe\,\textsc{vii}]7.81/[Fe\,\textsc{vii}]9.53 ratio was found to be the most robust tracer of electron density, due to its minimal sensitivity to $T_e$ across a broad density range ($10^4$–$10^8$~cm$^{-3}$) (see Fig. \ref{fig:Appen-density}). Using this ratio, we derived electron densities between $(5.20$–$7.88) \times 10^4$~cm$^{-3}$. These values are approximately an order of magnitude higher than those inferred from [Ne\,\textsc{v}] diagnostics, despite both ions having comparable ionization potentials (IP([Fe\,\textsc{vii}])=99.1, IP([Ne\,\textsc{v}])=97.1), suggesting [Fe\,\textsc{vii}] traces a denser, more compact ionized region closer to the nucleus.

While splitting into two components, the narrow component, associated with spatially extended emission, exhibited electron densities below $10^4$~cm$^{-3}$. In contrast, the outflow component displayed substantially higher electron densities, in the range of $(12.19$–$23.64) \times 10^4$~cm$^{-3}$. This trend matches that in [Fe\,\textsc{ii}], though [Fe\,\textsc{vii}] line ratios indicate densities several times higher, supporting the view that the outflow arises from denser, more turbulent regions near the active nucleus.

Table~\ref{tab:density} summarizes the density estimates derived from the $[$Fe\,\textsc{ii}$]$\,4.89/$[$Fe\,\textsc{ii}$]$\,5.34, $[$Ne\,\textsc{v}$]$\,8.91/$[$Ne\,\textsc{v}$]$\,24.3, and $[$Fe\,\textsc{vii}$]$\,7.81/$[$Fe\,\textsc{vii}$]$\,9.53 line ratios. 
A careful inspection of Fig.~\ref{fig:density} indicates that the observed ratios lie within the rising, density-sensitive portion of the diagnostic curves, with the exception of the narrow components, which approach the lower-density end. In contrast, the ratios associated with the outflow components occupy higher values, 
implying correspondingly higher electron densities.

\begin{figure}
    \centering
    \includegraphics[scale=0.44]{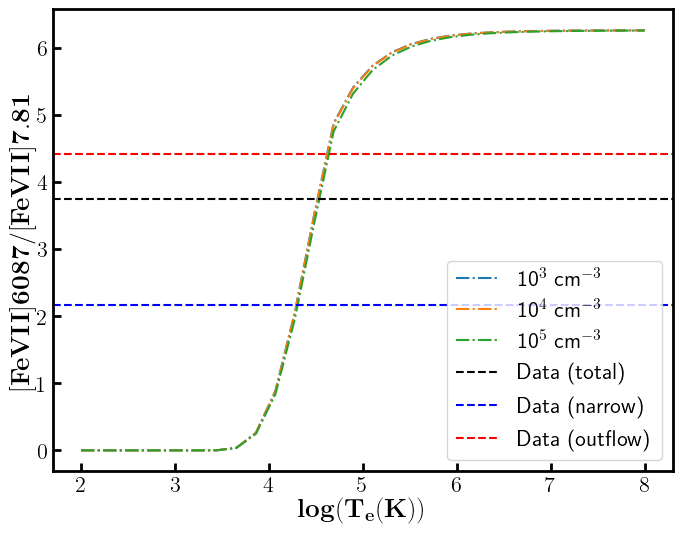}
    \caption{Variation of [Fe\,\textsc{vii}]$\lambda$6087~$\AA$ and [Fe\,\textsc{vii}]$\lambda$7.81~$\mu$m line ratio with electron temperature for different particular electron densities, which represent different curves in each plot and are denoted by individual legends in each plot. The horizontal line in the plot represents the observed line ratio.}
    \label{fig:temp}
\end{figure}

To estimate the $T_e$ of the ionised gas, we employed multiple [Fe\,\textsc{vii}] line ratios, focusing in particular on the [Fe\,\textsc{vii}]$\lambda$6087~$\AA$ and [Fe\,\textsc{vii}]$\lambda$7.81~$\mu$m ratio due to its high sensitivity to $T_e$ and relatively weak dependence on electron density ($n_e$), as illustrated in Figure~\ref{fig:temp}, compared to other two combinations shown in Fig. \ref{fig:Appen-temp}. We considered a range of plausible electron densities spanning $10^3$--$10^5$~cm$^{-3}$, adopting representative values of $10^3$, $10^4$, and $10^5$~cm$^{-3}$. The corresponding extinction-corrected temperature estimates for the integrated emission are $T_e = 32,911\pm 3,056$~K, $33,168\pm 3,072$~K, and $34,429\pm 3,159$~K, respectively.

When decomposing the emission into distinct kinematic components, we found a notable contrast in temperature between the narrow and broad components. For the narrow component, the derived temperatures are $T_e = 19,593\pm1,037 $~K, $19,709\pm 1,035$~K, and $20,300\pm 1,071$~K across the same density grid. In contrast, the broad component yields significantly elevated temperatures of $T_e = 41,658\pm 5,483$~K, $41,780\pm 5,681$~K, and $43,392\pm 6,571$~K. This substantial temperature difference may reflect additional heating mechanisms such as fast shocks or enhanced photoionisation in the broad-line emitting region, likely linked to AGN-driven outflows \citep{2021MNRAS.501L..54R}.

These [Fe\,\textsc{vii}]-based temperatures are considerably higher than those derived from classical low-ionization diagnostics, such as [O\,\textsc{iii}] and [N\,\textsc{ii}] line ratios, which typically yield $T_e \sim 15{,}000$--$18{,}000$~K \citep{2023ApJ...959..116N}. This discrepancy underscores the multi-phase nature of the ionised gas and highlights the importance of high-ionisation tracers in probing the energetic environment of the AGN narrow-line region.

These results reveal a clear stratification in both $n_e$ and $T_e$ across the ionised gas phases, with elevated values consistently associated with more highly ionised species and kinematically disturbed components. In particular, the [Fe\,\textsc{vii}] emission arises from regions of significantly higher density and temperature compared to those traced by lower ionisation lines, indicating its origin in more compact and energetically active zones \citep{2002ApJ...579..214R, 2005MNRAS.364.1041R}. This observed stratification is in agreement with theoretical expectations for AGN narrow-line regions shaped by a combination of photoionisation, radiation pressure, and shock excitation \citep{1997ApJS..110..287F, 2025FrASS..1248632R}.

\begin{table*}
    \centering
    \caption{Electron density measured from different tracers at different T$_e$}
    \label{tab:density}
    \begin{tabular}{lcccc}
    \hline
     Lines & n$_e$ (10000K) & n$_e$ (20000\,K) & n$_e$ (30000\,K) & n$_e$ (50000\,K)\\
       & (10$^4$ cm$^{-3}$) & (10$^4$ cm$^{-3}$) & (10$^4$ cm$^{-3}$) & (10$^4$ cm$^{-3}$) \\
     \hline
      $[Fe II]$4.89/$[Fe $II$]$5.34 (total)  & 0.50$\pm0.04$ & 0.65$\pm0.06$ & 0.71$\pm0.06$  & 0.82$\pm$0.06 \\
      $[Fe II]$4.89/$[Fe $II$]$5.34 (narrow)  & 0.27$\pm0.02$ & 0.34$\pm0.03$ & 0.37$\pm0.04$ & 0.43$\pm$0.04 \\
      $[Fe II]$4.89/$[Fe $II$]$5.34 (outflow)  & 2.58$\pm0.22$ & 3.34$\pm0.31$ & 3.65$\pm0.34$ & 4.25$\pm$0.04 \\
      $[Ne V]8.91/[Ne V]$24.3 (total) & 0.17$\pm 0.04$ & 0.33$\pm 0.09$ & 0.51$\pm0.10$  & 0.61$\pm$0.16\\
      $[Fe VII]7.81/[Fe VII]$9.53 (total) & 5.20$\pm2.25$ & 6.36$\pm3.04$ & 7.39$\pm3.74$ & 7.88$\pm$4.87 \\
      $[Fe VII]7.81/[Fe VII]$9.53 (narrow)$^{a}$ & -- & -- & -- & --\\
      $[Fe VII]7.81/[Fe VII]$9.53 (outflow) & 12.19$\pm3.05$ & 15.65$\pm4.11$ & 18.61$\pm5.17$ & 23.64$\pm$6.76 \\
      \hline
    \end{tabular}
    \vspace{2mm}
    \begin{minipage}{0.9\textwidth}
    \footnotesize
    \textbf{Note.} $^{a}$ This ratio lies below the theoretical limit, as shown in the right panel of Fig.~\ref{fig:density}.
    \end{minipage}
\end{table*}

\subsection{Ionized Gas Mass traced through H\,\textsc{i} Recombination Lines} \label{sec:M_ion}

We estimated the ionized gas mass ($M_{\rm ion}$) from the H\,\textsc{i} line luminosities using Equation~A5 of \citet{2022ApJ...930...14R}. All lines were corrected for extinction following the \citet{2023ApJ...950...86G} reddening law and adopting $A_V = 0.96$ \citep{2019MNRAS.486..691B, 2023ApJ...959..116N}. This method is generally applicable to any H\,\textsc{i} transition and is particularly effective when high-resolution optical or infrared spectra are available. The ionised mass is computed as:
\begin{equation}
    M_{\rm ion} = \left(\frac{L_{\rm H\,\textsc{i}}}{\mathrm{erg}~\mathrm{s}^{-1}}\right)
    \left(\frac{\epsilon_{\rm H\,\textsc{i}}}{\mathrm{erg}~\mathrm{cm}^{3}~\mathrm{s}^{-1}}\right)^{-1}
    \left(\frac{m_p^{\rm eff}}{n_p^{\rm eff}}\right),
    \label{eq:mass_HI}
\end{equation}
where $L_{\rm H\,\textsc{i}}$ is the luminosity of the H\,\textsc{i} recombination line, $\epsilon_{\rm H\,\textsc{i}}$ is the emissivity derived under Case~B recombination, $m_p^{\rm eff}$ is the effective proton mass, and $n_p^{\rm eff}$ is the effective proton density. 
Given the presence of metals and helium in the ionized gas, we adopted correction factors to account for the total particle mass and charge contribution: $m_p^{\rm eff} = 1.4\,m_p$ and $n_p^{\rm eff} = 1.1\,n_e$, consistent with assumptions in prior works \citep[e.g.,][]{2022ApJ...930...14R}.

We implemented this approach on the extinction-corrected fluxes of the near- and mid-infrared hydrogen recombination lines—Pa$\alpha$, Br$\gamma$, and Pf$\alpha$. The corresponding emissivities ($\epsilon_{\rm H\,\textsc{i}}$) were calculated using the \texttt{PyNeb} package \citep{2015A&A...573A..42L}, assuming Case~B recombination conditions as described in \citet{2006agna.book.....O}. We used an electron temperature of $T_e = 16{,}000$~K and an electron density of $n_e = 2000$~cm$^{-3}$ based on our previous estimate, and we derived ionized gas masses of $(1269 \pm 47)$~M$_\odot$ from Pa$\alpha$ and $(998 \pm 96)$~M$_\odot$ from Br$\gamma$. The close agreement between these estimates reinforces the consistency of our methodology across transitions that probe different depths of the ionised medium.

\subsection{Molecular Gas Excitation and Mass Estimation}


We investigated the physical conditions of the molecular gas in the nuclear region of NGC~4395 using both mid-infrared pure rotational and near-infrared ro-vibrational H$_2$ emission lines. In the mid-infrared, a total of 15 pure rotational transitions, from S(1) to S(15), were detected, providing sensitive diagnostics of the warm molecular gas and its excitation mechanisms \citep{1989ESASP.290..269S, 2005ApJ...633..105D, 2023A&A...675A..86K}. All line fluxes were corrected for extinction following \citet{2023ApJ...950...86G} and subsequently modelled using the PDR Toolbox \citep{2011ascl.soft02022P}.

\begin{figure}
    \centering
    \vbox{
    \hbox{
    \includegraphics[scale=0.43]{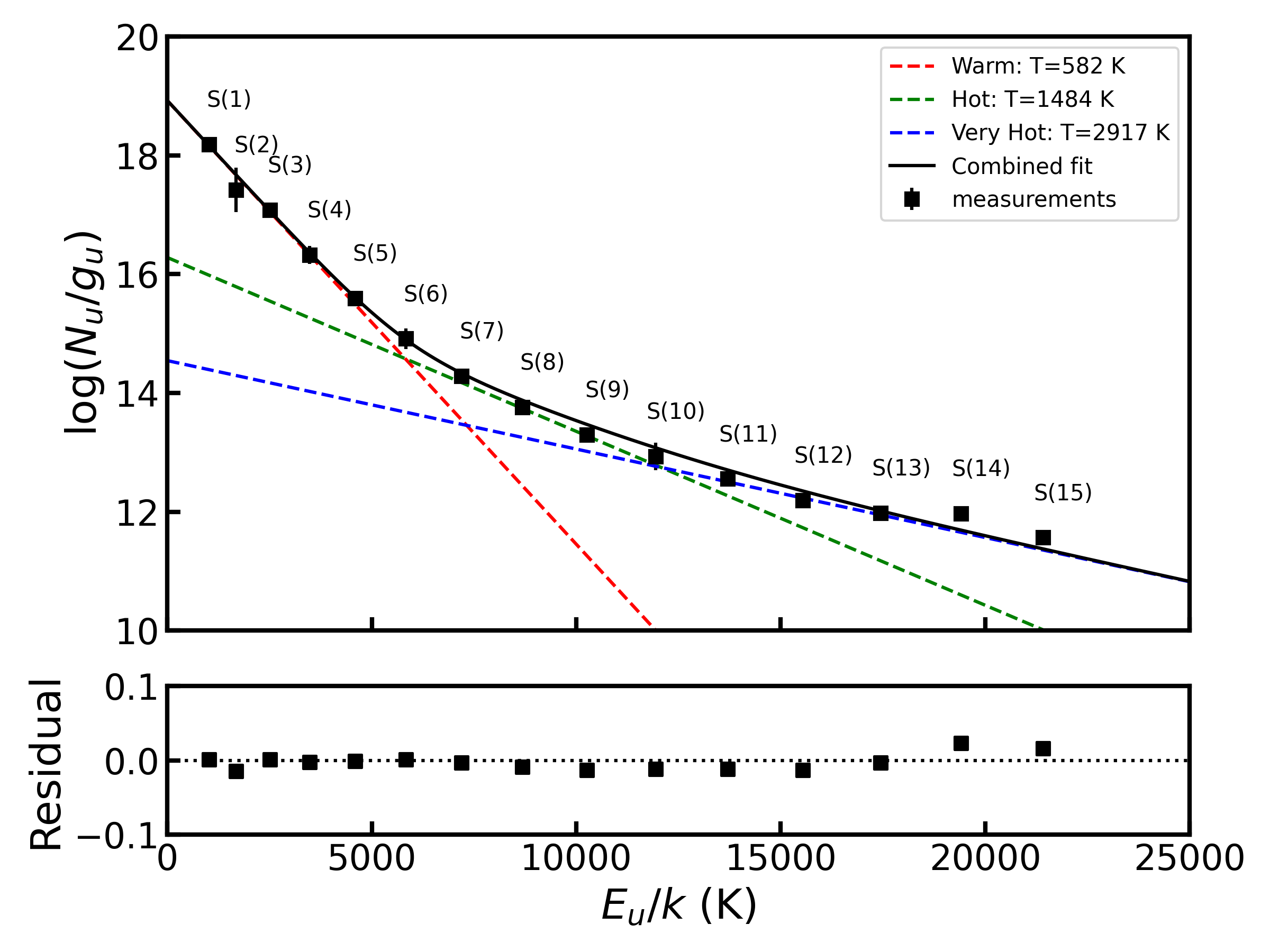}
    }
    \hbox{
    \includegraphics[scale=0.43]{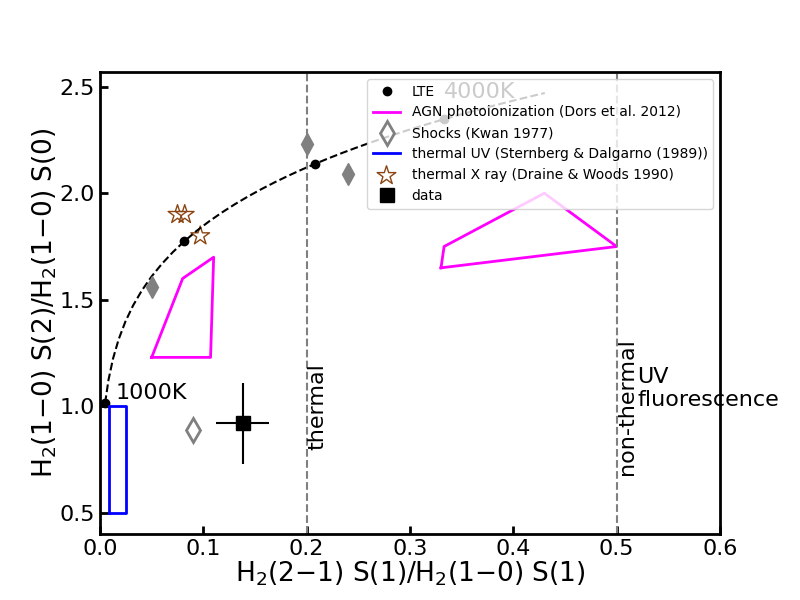}
    }
    }
    
    \caption{Upper panel: Rotational excitation diagram of the pure molecular hydrogen emission lines. The solid lines indicate the best-fit models, while the points represent the observed data. The residual panels show the normalized residuals, defined as (data$-$model)/data. 
    Lower panel: Lower panel: Excitation diagram of the H$_2$ ro-vibrational line ratios, H$_2$(2--1)~S(1)/H$_2$(1--0)~S(1) versus H$_2$(1--0)~S(2)/H$_2$(1--0)~S(0). Ratios with H$_2$(2--1)~S(1)/H$_2$(1--0)~S(1) $\lesssim 0.2$ indicate predominantly thermal excitation, while ratios $\gtrsim 0.5$ are characteristic of non-thermal excitation dominated by UV fluorescence \citep{1994ApJ...427..777M}. The black points connected by dotted lines represent LTE predictions for temperatures between 1000 and 4000~K \citep{2021MNRAS.506.2950R}. The pink shaded region corresponds to AGN photoionization \citep{2012MNRAS.422..252D}, the grey diamond symbols mark shock-excited regions \citep{1977ApJ...216..713K, 1995A&A...296..789S}, and the blue shaded region denotes thermal UV excitation \citep{1989ApJ...338..197S}. The brown star symbols indicate thermal X-ray excitation \citep{1990ApJ...363..464D}. The solid black square with error bars represents the measured line ratios for NGC~4395.
}
    \label{fig:H2_fit_combined}
\end{figure}

The excitation diagram constructed from the pure rotational lines (Upper panel of Fig.~\ref{fig:H2_fit_combined}) is best described by a three-component temperature model, which provides a significantly improved fit compared to a two-component solution. The best-fit temperatures correspond to a warm component at $582 \pm 1$~K, a hot component at $1484 \pm 6$~K, and a very hot component at $2917 \pm 169$~K. Such a temperature stratification is consistent with shock heating associated with AGN-driven winds or local turbulence for the warm and hot components, while the very hot component likely traces a small fraction of highly excited H$_2$ produced by non-thermal excitation processes, such as UV or X-ray pumping \citep[e.g.,][]{2010ApJ...724.1193O, 2004A&A...425..457R}.

To independently assess the excitation mechanism inferred from the mid-infrared analysis, we further examined the near-infrared ro-vibrational H$_2$ emission using extinction-corrected line ratios. Following the approach of \citet{2021MNRAS.506.2950R}, we constructed excitation diagrams involving diagnostic ratios such as H$_2$(2--1)~S(1)/H$_2$(1--0)~S(1) and H$_2$(1--0)~S(2)/H$_2$(1--0)~S(0), shown in the lower panel of Fig.~\ref{fig:H2_fit_combined}. These ratios are sensitive to deviations from local thermodynamic equilibrium and therefore provide a complementary probe of the excitation conditions. The observed line ratios are consistent with predominantly thermal excitation, in agreement with the temperature structure inferred from the pure rotational transitions.

The total line-of-sight molecular hydrogen column density derived from the PDR Toolbox model is
\[
N(\mathrm{H}_2)_{\rm total} = (1.20 \pm 0.07)\times 10^{20}\ \mathrm{cm^{-2}},
\]
dominated by the warm component $\left(1.21 \pm 0.07\right)\times 10^{20}\ \mathrm{cm^{-2}}$. The hot component contributes $\left(7.11 \pm 0.43\right)\times 10^{17}\ \mathrm{cm^{-2}}$, and the very hot phase contributes $\left(2.90 \pm 0.98\right)\times 10^{16}\ \mathrm{cm^{-2}}$.

Using a mean molecular weight of $\mu = 2.8$, and estimating the mass via $M_{\rm mol} = \mu m_p\, N_{\rm H_2}\, A$, we derive a total molecular gas mass of
\[
M_{\rm mol} = 936 \pm 52~M_\odot.
\]
The warm phase overwhelmingly dominates the mass budget ($930 \pm 52~M_\odot$), while the hot ($6 \pm 0.3~M_\odot$) and very hot ($0.2 \pm 0.07~M_\odot$) components contribute less than 1\%.
 
When we then further calculate the cold molecular gas mass traced through the CO(2--1) line in ALMA, we first derive the CO line luminosity using
\begin{equation}
L'_{\mathrm{CO}} = 3.25 \times 10^{7} \,
\left( \frac{S_{\mathrm{CO}}\Delta v}{\mathrm{Jy\,km\,s^{-1}}} \right)
\left( \frac{\nu_{\mathrm{obs}}}{\mathrm{GHz}} \right)^{-2}
\left( \frac{D_L}{\mathrm{Mpc}} \right)^{2}
(1+z)^{-3}.
\label{eq:co-mass}
\end{equation}
Since the observed transition is CO(2--1), we convert this luminosity to its CO(1--0) equivalent using a line ratio of 1.66 as found by other seyferts \citep{2021A&A...654A..24B}. The resulting CO(1--0) luminosity is then multiplied by the standard CO-to-H$_2$ conversion factor of $\alpha_{\mathrm{CO}} = (0.8-3.2)~M_\odot~(\mathrm{K\,km\,s^{-1}\,pc^2})^{-1}$ \citep{2021A&A...654A..24B} which strongly depends on the metalicity \citep{2012MNRAS.421.3127N, 2022A&A...663A..28S}, to obtain the total cold molecular gas mass. Following this method, we find that the total cold molecular gas has a mass of $(0.95-3.2) \times10^{6}~M_\odot$.

So, this warm/hot molecular gas mass ($>$500K) to cold molecular gas mass ($<$100K) is  ratio 2.4--9.8)$\times10^{-4}$, but hot molecular ($>$1000K) i.e. considering the hot and very hot component to cold molecular gas ratio (1.6--6.5)$\times10^{-6}$ which is in the range that found in other seyferts \citep{2020MNRAS.496.4857R}

Then we calculated the warm/hot molecular gas($>$500K) mass-to-ionized gas mass ratio, $M_{\mathrm{H}_2} / M_{\rm ion}$, and found that is 0.7--0.9. This is consistent with values reported in the literature, for instance, in NGC~7319, \cite{2022A&A...665L..11P} observed a similar ratio, indicative of active feedback processes where the ionised phase can rival or exceed the warm/hot molecular mass content. 


\subsection{Excitation Mechanisms of the PAH and H$_2$ Features} \label{sec:PAH_excitation}


We further investigated the excitation mechanism responsible for the detected PAH features to determine whether the observed emission is primarily associated with star formation or modified by AGN-related processes. From the measured PAH bands, we derive intensity ratios of PAH$_{3.3}$/PAH$_{11.3} = 0.18 \pm 0.08$ and PAH$_{6.2}$/PAH$_{11.3} = 0.97 \pm 0.07$. These relatively low ratios indicate a PAH population that is predominantly neutral and biased toward larger molecular sizes ($\gtrsim 200$-carbon) \citep{10.1093/mnras/stae1535}. Such conditions are consistent with the survival of robust, less-ionized grains in environments exposed to a hard radiation field, where small PAH molecules are preferentially destroyed or ionized. 

The 7.7 and 12.7~$\mu$m features are typically associated with smaller and more ionized PAH molecules, and their absence indicates that the emission is dominated by large, neutral PAHs that can withstand the harsher AGN radiation field. 

Recent spatially resolved \textit{JWST} studies have reported similar behavior in several nearby active galaxies. The nuclear regions of Seyferts and LINERs often exhibit a suppression of the shorter-wavelength (ionized) PAH bands and an enhancement of the 11.3~$\mu$m feature, indicative of a higher neutral PAH fraction \citep[e.g.,][]{2023ApJ...944L...7S, 2024A&A...691A.162G, 10.1093/mnras/stae1535}. Such trends are interpreted as the result of PAH processing by intense radiation fields or shocks, leading to the preferential survival of large neutral PAH molecules in edges of torus. 

We investigated the excitation mechanism of the warm molecular gas through mid-infrared H$_2$–to–PAH diagnostics. The H$_2$\,S(3)\,(0--0)/PAH$_{11.3}$ ratio was found to be $-0.03$ in logarithmic scale, placing the source firmly within the AGN-ionized regime. This suggests that the H$_2$ emission is primarily powered by shock excitation rather than UV heating from star-forming regions. Similar elevated H$_2$/PAH ratios have been reported in nearby radio-loud Seyferts such as 3C\,293 and NGC~3884, where AGN jets and outflows interact strongly with the surrounding interstellar medium \citep{10.1093/mnras/stz1316, 2024A&A...691A.162G, 2025ApJ...982...69R, 2025ApJ...993..217D}. 

We further evaluated the [Fe\,\textsc{ii}]\,5.34\,$\mu$m/PAH$_{11.3}$ ratio, a well-known tracer of shocks in partially ionized gas. We obtained a value of $-0.6$ (log scale), which is comparable to those measured in the nuclei of 3C\,293 and CGCG~012--070 \citep{2025ApJ...982...69R}. Elevated [Fe\,\textsc{ii}]/PAH ratios typically signal enhanced mechanical heating, where shocks release Fe from dust grains, leading to stronger [Fe\,\textsc{ii}] emission relative to PAH features. Similar trends have been reported in various Seyfert and LINER galaxies \citep[e.g.,][]{2019MNRAS.487.1823L}, reinforcing the association between AGN-driven feedback and shock excitation of the ISM.

Taken together, the combination of low PAH$_{6.2}$/PAH$_{11.3}$ and PAH$_{3.3}$/PAH$_{11.3}$ ratios, the absence of the 7.7 and 12.7~$\mu$m features, and the elevated H$_2$/PAH and [Fe\,\textsc{ii}]/PAH ratios point to a scenario where the PAH-emitting regions are dominated by shocks associated with AGN activity rather than photoionization from ongoing star formation. This interpretation is consistent with other jet--ISM interaction systems, where mechanical feedback from AGN outflows enhances molecular hydrogen emission while simultaneously suppressing or destroying the smallest PAH carriers. Such shock-dominated conditions illustrate the significant impact of AGN feedback on the excitation and chemical state of the circumnuclear ISM, even in systems with modest radiative luminosities.

\section{Discussion} \label{sec:discussion}

\subsection{Multiphase Outflows}

\begin{figure*}[ht]
    \centering
    \vbox{
    \hbox{
    \includegraphics[scale=0.25]{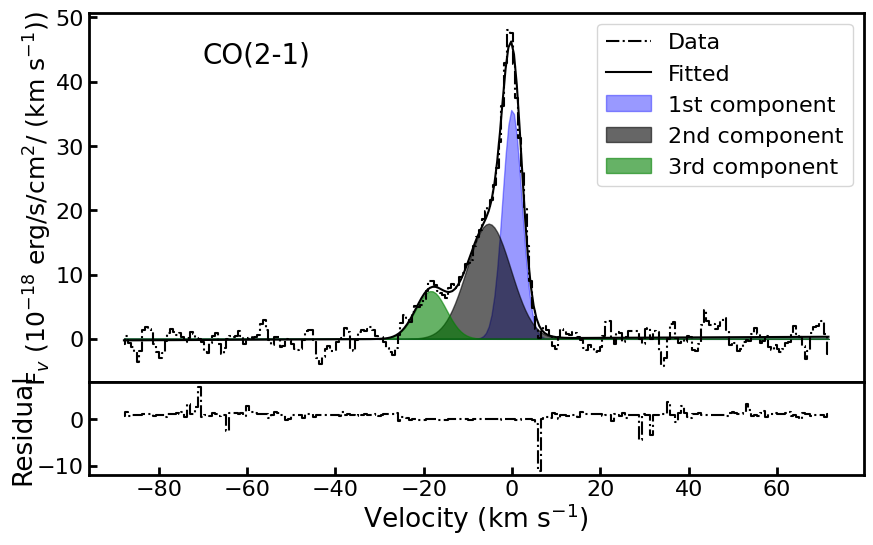}
    \includegraphics[scale=0.25]{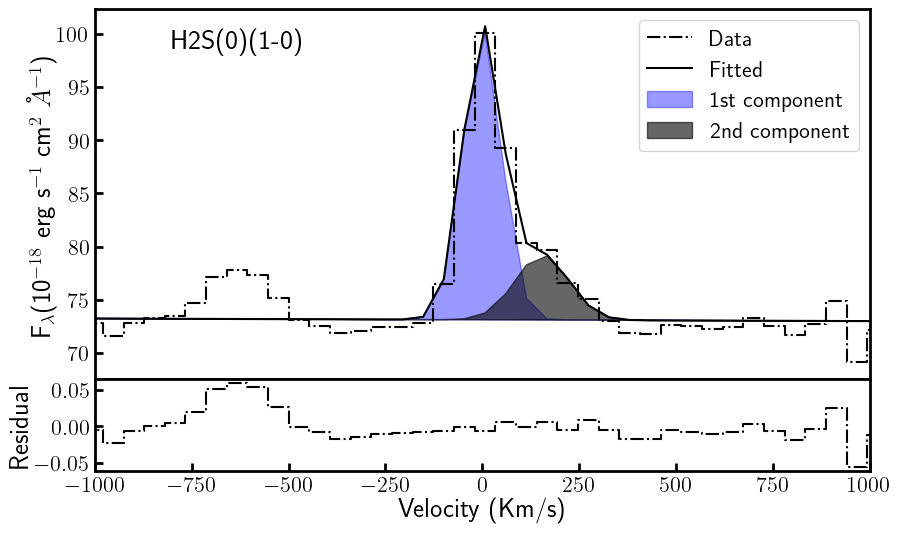}
    \includegraphics[scale=0.25]{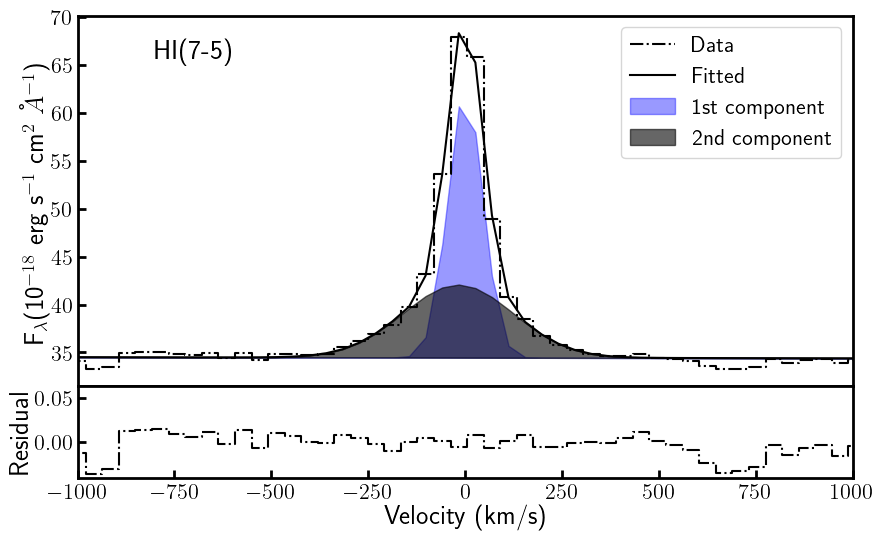}
    }
    \hbox{
    \includegraphics[scale=0.25]{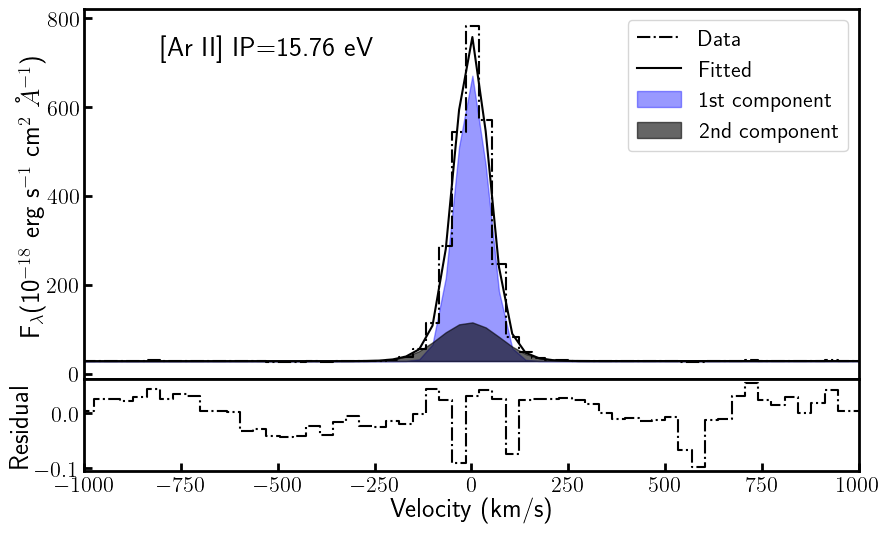}
    \includegraphics[scale=0.25]{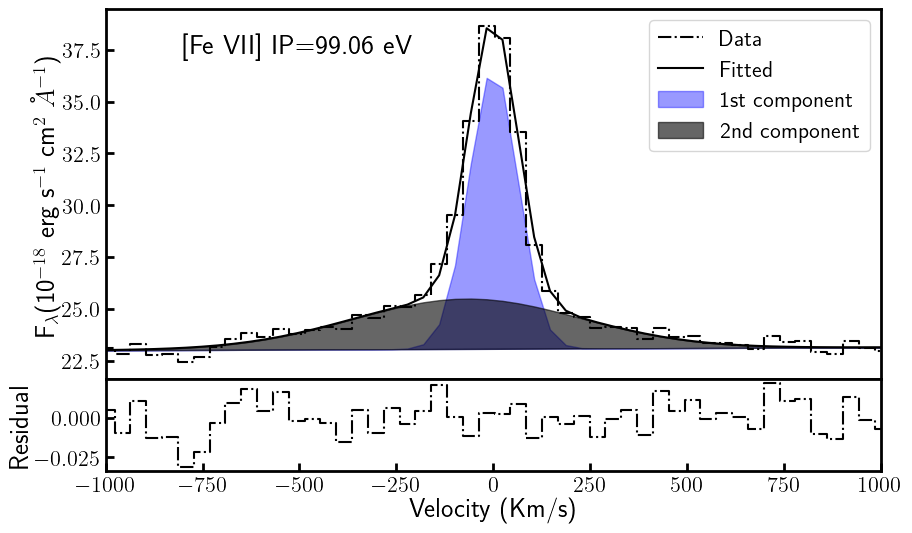}
    \includegraphics[scale=0.25]{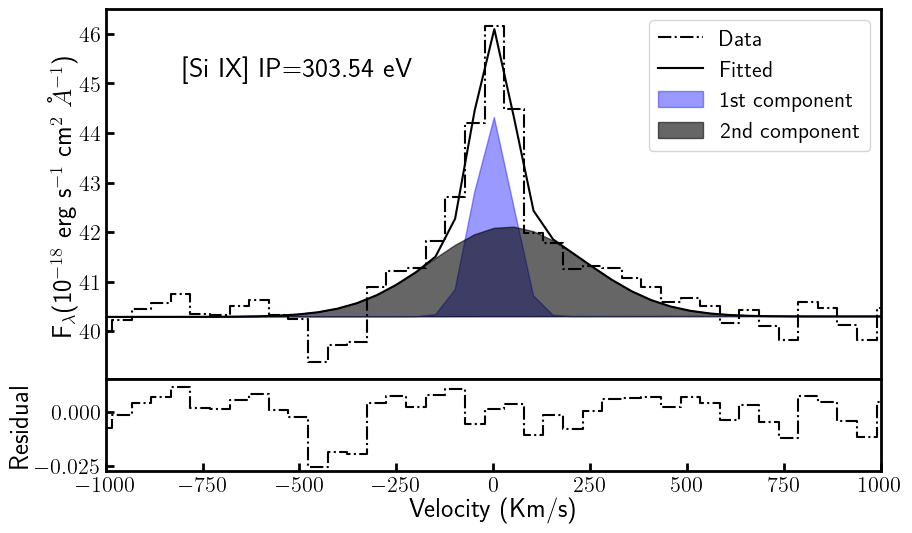}
    }
    }
    \caption{Visualisation of outflows across multiple ionised gas phases: top panels show cold molecular gas (top left), hot molecular gas (top middle), and neutral HI gas (top right). The bottom panel illustrates ionised outflows traced by different elements spanning a range of ionisation states, from low (left) to highly ionised gas (right). Corresponding spectral lines and IP are indicated in each panel.}
    \label{fig:outflow-detec}
\end{figure*}

Our spectral decomposition analysis revealed that several fine-structure and molecular emission lines required more than a single Gaussian component to achieve statistically robust fits. This additional broadened and velocity-shifted component is indicative of a kinematically distinct gas population consistent with outflowing material. The detection of these features across a broad range of ionisation states and gas temperatures strongly supports the presence of multiphase outflows driven by AGN activity in NGC~4395 which is shown in Fig. \ref{fig:outflow-detec}.

\subsubsection{Ionized Gas Outflows}

Clear signatures of outflows are observed across a broad range of ionised gas tracers in our data, spanning from low to high-ionisation potential species. The most prominent evidence for fast nuclear outflows comes from coronal lines such as [Fe\,\textsc{vii}], [Mg\,\textsc{viii}] and [Si\,\textsc{ix}], which are formed in gas with ionisation potentials exceeding 100~eV. These lines display broad, asymmetric profiles and require the addition of secondary kinematic components with velocity shifts of several hundred km\,s$^{-1}$ relative to systemic. Such features are characteristic of fast, highly ionised winds originating from the inner regions of the AGN, potentially close to the accretion disk or the base of the ionisation cone \citep{2002ApJ...579..214R, 2011ApJ...739...69M, 2025MNRAS.538.2800R}.

In addition to the prominent high-ionisation coronal lines, we identified broadened and blueshifted components in several low- to intermediate-ionisation forbidden lines, including [Ar\,\textsc{ii}] (ionisation potential $\sim$15.8~eV), [Ne\,\textsc{ii}] ($\sim$21.6~eV), and [Ca\,\textsc{iv}] ($\sim$50.9~eV). These features point to the presence of ionised outflows that extend well beyond the innermost coronal region, impacting the NLR and potentially influencing the larger-scale ISM of the host galaxy. 

The observed kinematic signatures in low-ionisation lines are particularly noteworthy, as they offer a crucial window into the more extended and less extreme regions of the outflow. These lines are less affected by the extreme ionising conditions near the nucleus, and their detection implies that the AGN-driven winds are capable of entraining and accelerating gas located farther from the central engine. Such findings are consistent with MIR IFU studies that demonstrate spatially resolved outflows extending over hundreds of parsecs to kiloparsec scales in nearby Seyferts and quasars \citep{2022A&A...665L..11P, 2023A&A...672A.108A, 2024ApJ...974..195Z}.

Furthermore, we detected outflow components in [Fe\,\textsc{ii}] lines, which are known to be enhanced in the presence of shocks due to their efficient excitation via collisional processes in partially ionised zones \citep{2001A&A...369L...5O, 2024ApJ...960...41M, 2009MNRAS.394.1148S}. The presence of these features suggests a possible contribution from shock-induced ionisation in addition to photoionisation by the AGN, reinforcing a multi-phase picture of AGN feedback that involves both radiative and mechanical drivers. 

This diversity in ionisation states and spatial extents implies a stratified, multi-phase outflow structure, where the acceleration and ionisation conditions vary with gas density, composition, and distance from the AGN. Gas with differing ionisation states and densities responds variably to the AGN radiation field and mechanical forces \citep{2022ApJ...924...82F}.
Similar stratification has been reported in nearby Seyfert galaxies and luminous quasars \citep{2013ApJ...768...75R, 2021MNRAS.506.2950R, 2023ApJ...942L..37A, 2025arXiv250321921M}, where compact, fast outflows in highly ionised lines coexist with more extended, slower flows traced by lower-ionisation species. Our findings reinforce this scenario, showing that outflows traced by [Ne\,\textsc{ii}], [Mg\,\textsc{vii}], [Si\,\textsc{ix}] and other MIR lines can effectively probe the kinematics of ionised gas phases otherwise inaccessible due to optical extinction.

To characterise the kinematics of the ionised outflowing gas, we computed the outflow velocity ($V_{\mathrm{out}}$) for each fine-structure line with outflows using the standard formalism:
$ V_{\mathrm{out}} = V_{\mathrm{shift}} + 2\sigma_{\mathrm{out}}$ ,
where $V_{\mathrm{shift}}$ is the velocity offset between the broad (outflowing) and the narrow (systemic) components, and $\sigma_{\mathrm{out}}$ is the velocity dispersion of the broad component (corrected for instrumental effects). This definition accounts for both the bulk motion and the internal turbulent broadening of the outflow and is widely adopted in studies of AGN-driven winds \citep[e.g.,][]{2025ApJ...984...20N, 2025A&A...695A...6P}.

\begin{figure}
    \centering
    \hspace*{-0.5cm}\includegraphics[scale=0.46]{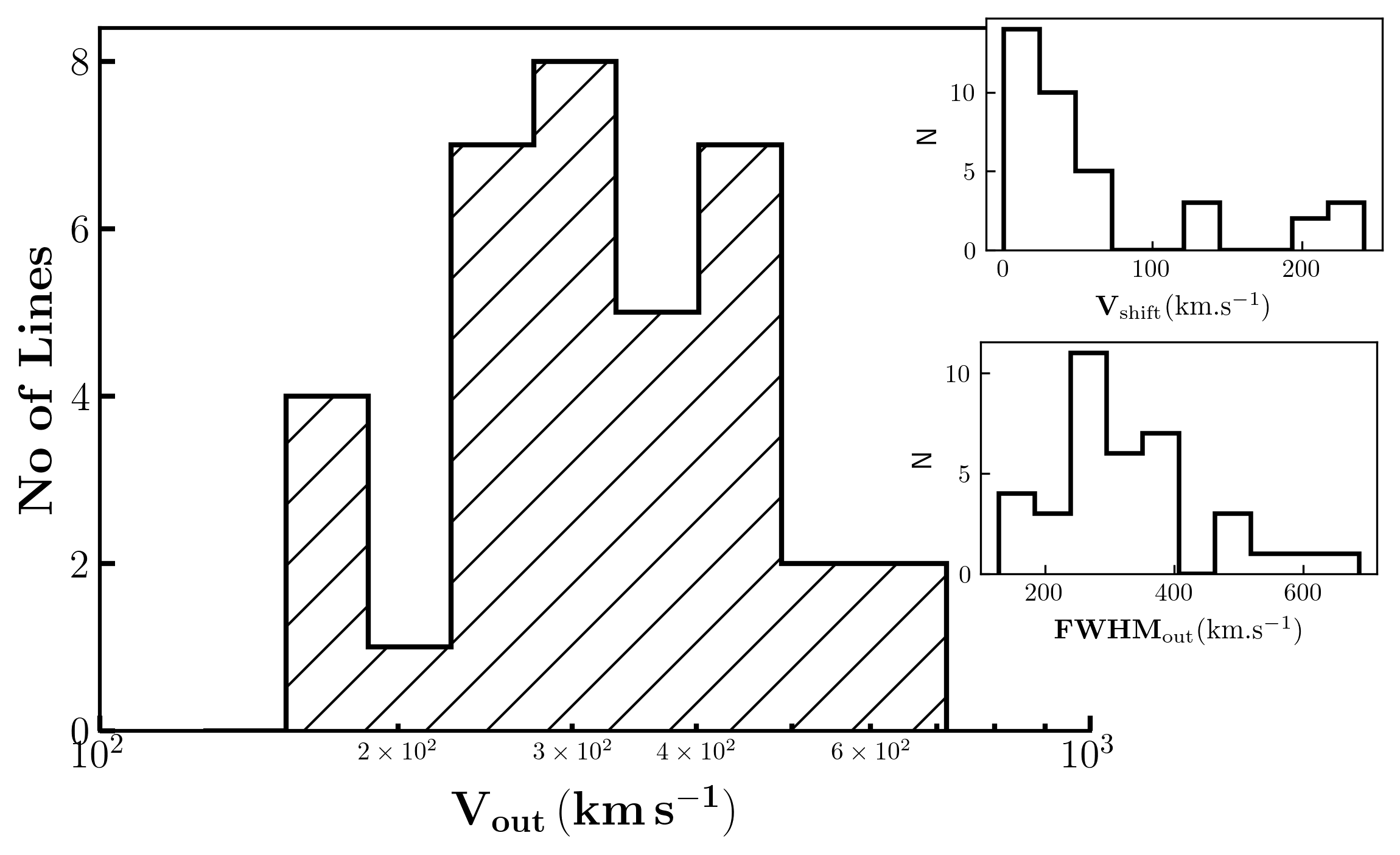}
    \caption{Distribution of the outflow velocity, V$_{\mathrm{out}}$ (V$_{\mathrm{out}}$ = V$_{\mathrm{shift}}$ + 2$\sigma_{\mathrm{out}}$), of ionized gas traced by fine-structure lines. The insets show the velocity shift \textbf{(V$_{\mathrm{shift}}$)} and the FWHM (FWHM$_{out}=2.355\times \sigma_{out}$) of the outflowing components.}
    \label{fig:v_out}
\end{figure}

The computed $V_{\mathrm{out}}$ range between 127 and 716~km\,s$^{-1}$, with a median value of 318~km\,s$^{-1}$. The distribution is illustrated in Figure~\ref{fig:v_out}. Notably, a subset of lines shows elevated outflow velocities ($V_{\mathrm{out}} > 400$~km\,s$^{-1}$), predominantly associated with shock-excited [Fe\,\textsc{ii}] emission and high-ionisation coronal lines such as [Fe\,\textsc{vii}], [Mg\,\textsc{vii}] and [Si\,\textsc{ix}]. These lines are expected to trace gas located closer to the active nucleus, where stronger radiation fields and mechanical feedback mechanisms, such as radiation pressure or jet-driven shocks, are capable of imparting higher momentum, resulting in faster outflowing material.

The observed velocities are consistent with typical values found in warm ionised outflows in nearby Seyfert galaxies \citep{2018MNRAS.476.2760F, 2021MNRAS.507...74R, 2024MNRAS.530.3059G, 2025ApJ...984...20N}, though they remain significantly below the velocities observed in UltraFast Outflows detected in X-ray spectra, which can exceed 10,000~km\,s$^{-1}$ \citep{2021NatAs...5...13L}. Together with the detection of lines spanning a wide range of ionisation potentials, this supports the presence of a stratified outflow structure, where different gas phases are accelerated to varying degrees depending on their density, ionisation state, and distance from the central engine.

We estimated the mass of the ionized outflowing gas using the H\,\textsc{i} recombination lines and the formalism described in Section~\ref{sec:M_ion} (Equation~\ref{eq:mass_HI}), substituting the line luminosity with that of the broad outflowing H\,\textsc{i} component. The derived outflow masses are $M_{\mathrm{out}} = 503 \pm 35~M_{\odot}$ and $370 \pm 82~M_{\odot}$ from the Pa$\alpha$ and Br$\gamma$ lines, respectively. The corresponding mass outflow rates \footnote{These are lower limit, as we considered the outflow radius as 0.5$^{\prime\prime}$ which is mostly upper limit}, computed as $\dot{M}_{\mathrm{out}} = M_{\mathrm{out}}\,V_{\mathrm{out}}/R_{\mathrm{out}}$, are $0.03~M_{\odot}\,\mathrm{yr^{-1}}$ and $0.01~M_{\odot}\,\mathrm{yr^{-1}}$ for Pa$\alpha$ and Br$\gamma$, respectively, adopting an outflow radius of $R_{\mathrm{out}} = 10.5$~pc (0$\farcs$5), which corresponds to the size of our spectral extraction aperture.

We further derived the ionized outflow mass using the high-ionization [Ne\,\textsc{v}]\,14.32\,$\mu$m line, adopting $n_{\mathrm{e}} = 10^{4}$~cm$^{-3}$ and $T_{\mathrm{e}} = 1.5 \times 10^{4}$~K, and following Equation~B5 of \citet{2024ApJ...974..195Z}. In this calculation, we employed Pa$\alpha$ instead of Pf$\alpha$ (used in \citealt{2024ApJ...974..195Z}), since Pa$\alpha$ is the strongest H\,\textsc{i} transition in our line list. This yields $M_{\mathrm{out}} = 175 \pm 5~M_{\odot}$.

Applying the same methodology as \citet{2024ApJ...974..195Z}, and adopting identical assumptions,\footnote{The emissivities of the relevant transitions vary by less than 1\% across densities of $10^{3}$--$10^{4}$~cm$^{-3}$. These emissivities were obtained from the CHIANTI database: \url{https://www.chiantidatabase.org/chianti_chiantipy.html}.} We likewise estimated the outflowing mass traced by the high-ionization [Fe\,\textsc{vii}]\,7.81\,$\mu$m and [Mg\,\textsc{vii}]\,3.03\,$\mu$m lines, both of which show clear signatures of outflowing kinematics. Using the respective emissivities, we derived $M_{\mathrm{out}} = 112 \pm 40~M_{\odot}$ from [Fe\,\textsc{vii}] and $M_{\mathrm{out}} = 178 \pm 21~M_{\odot}$ from [Mg\,\textsc{vii}]. These values indicate that the total coronal-line outflowing mass lies in the range $\sim$110--180~$M_{\odot}$, with consistent estimates across all three high-ionization tracers. The corresponding mass outflow rate, $\dot{M}_{\mathrm{out}}$, is 0.002--0.006~$M_{\odot}\,\mathrm{yr^{-1}}$.

The associated kinetic power of the outflow was calculated as  
\begin{equation}
    \dot{E}_{\mathrm{out}} = \frac{1}{2}\,\frac{M_{\mathrm{out}}}{R_{\mathrm{out}}}\,V_{\mathrm{out}}^{3}~{\rm erg~s^{-1}},
\end{equation}
where we adopted $R_{\mathrm{out}} = 10.5$~pc (0$\farcs$5), corresponding to the size of our aperture.  
This gives $\dot{E}_{\mathrm{out}} = (2.71 \pm 0.19)\times10^{39}$~erg~s$^{-1}$ for Pa$\alpha$ and $(6.74 \pm 1.48)\times10^{38}$~erg~s$^{-1}$ for Br$\gamma$.  

Similarly, we derived $\dot{E}_{\mathrm{out}} = (6.38 \pm 1.38)\times10^{36}$~erg~s$^{-1}$ from [Ne\,\textsc{v}], $(1.90 \pm 0.78)\times10^{38}$~erg~s$^{-1}$ from [Fe\,\textsc{vii}], and $(2.41 \pm 0.38)\times10^{38}$~erg~s$^{-1}$ from [Mg\,\textsc{vii}]. The outflow energetics inferred from [Fe\,\textsc{vii}] and [Mg\,\textsc{vii}] are over an order of magnitude higher than those derived from [Ne\,\textsc{v}], yet about an order of magnitude lower than those from the low-ionization H\,\textsc{i} lines. This suggests that [Fe\,\textsc{vii}], which traces a more powerful outflow, likely originates closer to the central engine than [Ne\,\textsc{v}], as it is associated with denser gas (see Section~\ref{sec:ne-Te}). In contrast, the lower-ionization outflows, characterized by lower densities, likely arise farther from the nucleus where the gas interacts more strongly with the surrounding ISM.  

In all cases, the kinetic power of the outflow remains several orders of magnitude below the AGN bolometric luminosity, $(1.95$--$4.97)\times10^{41}$~erg~s$^{-1}$, and also lower than the estimated jet power, $(1.3 \pm 0.3)\times10^{40}$~erg~s$^{-1}$ \citep{2023ApJ...959..116N}. The corresponding kinetic coupling efficiencies, defined as $\dot{E}_{\mathrm{out}}/L_{\rm bol}$, range between 0.4--1.4\% for the low-ionization H\,\textsc{i} lines and 0.003--0.12\% for the coronal-line outflows, assuming the lower limit of $L_{\mathrm{bol}} = 1.95\times10^{41}$~erg~s$^{-1}$.  

These efficiencies are consistent with the theoretical threshold for H\,\textsc{I} outflows but fall well below the $\sim$1\% level expected for the coronal-line outflows to drive significant AGN feedback capable of regulating star formation \citep{2018NatAs...2..198H}. This suggests that the low-ionization outflows can more effectively influence the surrounding interstellar medium of the host galaxy than the coronal-line outflows.

\subsubsection{Outflows in Molecular Gas}

Beyond the ionised gas phases, we also found evidence of outflows in the warm and hot molecular gas components. A prominent example is the H$_2$(1--0)~S(0) ro-vibrational line at 2.2245\,$\mu$m, which exhibits a secondary broadened kinematic component with a velocity shift of $\sim$154~km~s$^{-1}$ relative to the narrow component and a FWHM of 163~km~s$^{-1}$. This line is known to trace hot ($T \sim$ a few 1000~K) molecular gas, typically excited by shocks or intense radiation fields in AGN environments \citep[e.g.,][]{2014MNRAS.439.2701H, 2018MNRAS.478.3100R}. The detection of this broadened component supports the presence of AGN-driven molecular outflows in the nuclear region of the galaxy.

In addition to this ro-vibrational line, signatures of molecular outflows were also found in several pure rotational H$_2$ transitions observed in the mid-infrared. For warm molecular gas, we detected outflow components in the H$_2$~S(2), S(3), S(4), and S(5) (0--0) lines. These exhibit low velocity offsets ($<$10~km~s$^{-1}$) but are broadened, with FWHM values of 121, 146, 111, and 106~km~s$^{-1}$, respectively. Furthermore, we identified an outflow component in the H$_2$(0--0)~S(8) line at 5.0353\,$\mu$m, which shows a velocity shift of $\sim$10~km~s$^{-1}$ and a broader profile with FWHM $\sim$180~km~s$^{-1}$, indicative of a hotter molecular gas phase. These findings are consistent with previous observations of warm molecular outflows in AGN \citep[e.g.,][]{2016A&A...588A..46M, 2022A&A...665L..11P, 2024ApJ...974..127C, 2025ApJ...982...69R}, reinforcing the multi-phase nature of AGN feedback and its influence on the host galaxy's molecular interstellar medium.


We detected outflows in several H$_2$ transitions, including the pure rotational S(2)--S(5) (0--0) lines and the ro-vibrational H$_2$(1--0)~S(0) line at 2.22~µm, but not in the stronger H$_2$(1--0)~S(1) line at 2.12~µm. This differential detection reflected their distinct excitation conditions (upper level excitation energy for H$_2$(1--0)~S(0) is 6471K and H$_2$(1--0)~S(1) is 6956K \citep{1984CaJPh..62.1639D}).
The H$_2$(1--0)~S(1) line is often associated with warm molecular gas that can be excited by UV fluorescence, shocks, or X-ray heating \citep{1987ApJ...322..412B,1996ApJ...466..561M}. However, for this source, the measured ro-vibrational line ratios (see lower panel of Fig. \ref{fig:H2_fit_combined}) place the emission firmly in the \emph{thermal excitation region} of the diagnostic diagram of \citet{1994ApJ...427..777M}, rather than in the non-thermal UV-fluorescent regime. This indicates that the dominant excitation mechanism is thermal, collisional heating, most plausibly driven by shocks associated with the outflow interacting with the surrounding ISM, rather than radiative UV pumping.

We estimated the warm/hot molecular outflow mass by computing the fraction of outflowing flux ($f$) relative to the total flux of the ro-vibrational H$_2$ (1--0)~S(0) line, obtaining $f = 0.24$. This yields an outflowing molecular mass of $f \times M{\rm mol} = 223~M_\odot$. Using the outflow velocity derived from the H$_2$(1--0)~S(0) outflow component (223 km s$^{-1}$), we infer a mass outflow rate of $0.005~M\odot,\mathrm{yr^{-1}}$. The corresponding kinetic power, $\dot{E}_{\mathrm{out}}$, is $1.07 \times 10^{38}~\mathrm{erg,s^{-1}}$, consistent with the value obtained from the coronal-line outflow.


In addition to the ionised, hot and warm molecular phases, signatures of outflows are also evident in the cold molecular gas component, as traced by the CO(2--1) rotational transition at 1.3~mm observed with \textit{ALMA}. The CO(2--1) line profile reveals a complex kinematic structure composed of three distinct components: a narrow, systemic component, and two broader components blueshifted with respect to the systemic velocity, indicative of cold molecular outflows as shown in upper left panel of Fig. \ref{fig:outflow-detec}.

The total integrated flux of the CO(2--1) line is measured to be $4.84 \times 10^{-16}$~erg~s$^{-1}$~cm$^{-2}$. The line profile reveals a narrow systemic component with a flux of $1.95 \times 10^{-16}$~erg~s$^{-1}$~cm$^{-2}$ and a FWHM of 5.1~km~s$^{-1}$, consistent with dynamically cold, gravitationally bound gas. In addition, two outflowing components are detected, with fluxes of $2.23 \times 10^{-16}$ and $0.65 \times 10^{-16}$~erg~s$^{-1}$~cm$^{-2}$, blueshifted by 5.3~km~s$^{-1}$ and 18.4~km~s$^{-1}$ relative to the systemic component, and FWHMs of 11.71~km~s$^{-1}$ and 8.2~km~s$^{-1}$, respectively. The detection of multiple velocity components in the CO(2--1) line suggests a stratified or multi-phase outflow structure, potentially tracing distinct spatial zones or evolutionary stages of gas entrainment and acceleration. Similar structured outflows in the cold molecular phase have been observed in other AGN-hosting systems \citep[e.g.,][]{2019MNRAS.483.4586F,2023MNRAS.522.3753D}, underscoring the impact of AGN-driven winds on even the densest and coldest regions of the interstellar medium.  

These relatively narrow line widths suggest that the cold molecular gas traced by CO(2--1) is less turbulent than the warm or hot molecular components. The kinematics of this cold gas differ substantially from those of the warm/hot molecular outflow traced by the H$_2$ lines. Moreover, the cold CO-emitting gas is displaced by $\sim$20~pc from the nucleus, in a direction perpendicular to the AGN jet towards the north direction \citep{2023ApJ...959..116N}, whereas the warm and hot molecular gas is concentrated in the central region coincident with the jet. This spatial and kinematic decoupling implies that the warm/hot molecular component may not share the same origin as the cold molecular gas.

Using the CO(2--1) line and Equation~\ref{eq:co-mass}, together with the prescriptions outlined in the previous section, we estimated the mass of the outflowing cold molecular gas from the velocity–integrated fluxes of the two outflow components. The total cold molecular outflow mass is $(0.57$--$2.27)\times10^{6}\,M_\odot$, with the first component contributing $(0.44$--$1.76)\times10^{6}\,M_\odot$ and the higher–velocity component contributing $(0.13$--$0.51)\times10^{6}\,M_\odot$. The corresponding mass outflow rates are $(0.31$--$1.25)\,M_\odot\,\mathrm{yr^{-1}}$ for the first component and $(0.15$--$0.61)\,M_\odot\,\mathrm{yr^{-1}}$ for the second. These values are significantly higher than the outflow rates derived for the warm/hot molecular and ionised gas phases. The associated kinetic powers are $(2.08$--$8.33)\times10^{37}\,\mathrm{erg\,s^{-1}}$ and $(2.85$--$11.4)\times10^{37}\,\mathrm{erg\,s^{-1}}$ for the first and second components, respectively.

\vspace{1cm}

By comparing the outflow signatures across different gas phases and
their velocity shifts relative to the narrow disk component, the overall
pattern reveals a stratified, multiphase outflow shaped by
phase-dependent obscuration. Most fine-structure ionised lines, along
with the intermediate-ionisation coronal lines [Fe\,VII] and [Si\,VII],
exhibit blueshifted components, consistent with emission from an
unobscured near-side outflow. In contrast, the very high-ionisation
coronal line [Si\,IX] and the H$_2$ molecular lines are redshifted,
implying that their near-side emission is attenuated by nuclear dust or
the inner torus, leaving the receding component dominant. Such
ionisation-dependent velocity behaviour indicates that different species
originate at different radii and encounter varying levels of extinction,
consistent with a biconical outflow geometry---already spatially
resolved for the [O\,III] outflow using \textit{HST} by
\citet{2023ApJ...959..116N}. Similar stratified and obscured biconical
structures have been reported in other AGN
\citep[e.g.][]{Crenshaw_2010, 2013ApJS..209....1F,
2011ApJ...739...69M}, supporting a scenario in
which the observed kinematics of each gas phase depend on both its
location within the outflow and the line-of-sight extinction.

\subsection{Ionization Potential Dependence of Outflow Properties}
\label{sec:ep-outflow}

To probe the physical and ionisation structure of the ionised outflows, we explored the relationship between the kinematic properties of the emitting gas and the IP of the corresponding ionic species. The IP serves as a proxy for the ionisation state and can reveal the stratification in both energy and spatial origin within the outflowing medium \citep[e.g.,][]{2002ApJ...579..214R, 2014A&A...562A..21C}.

\begin{figure}
    \centering
    \vbox{
    \includegraphics[scale=0.45]{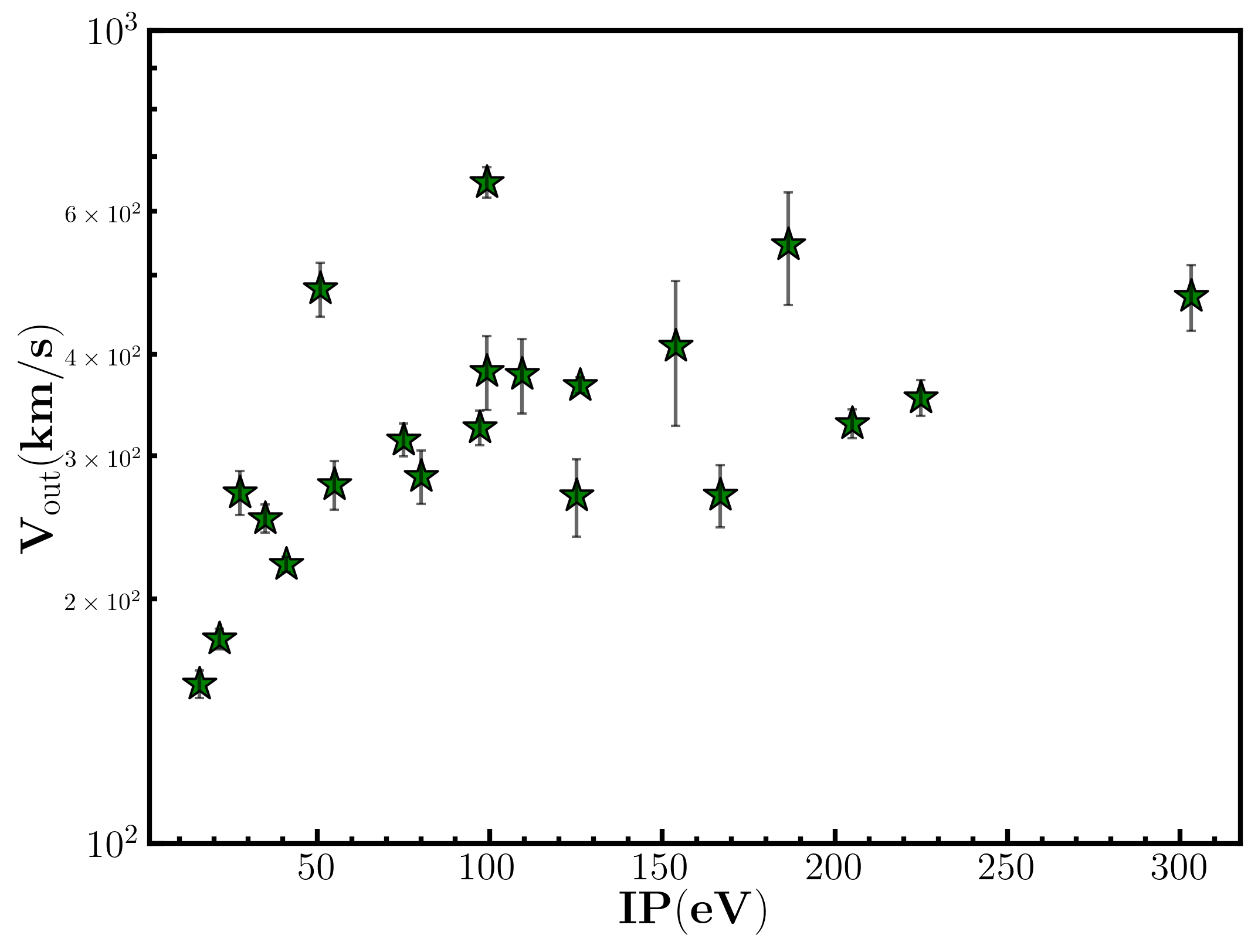}
    \includegraphics[scale=0.45]{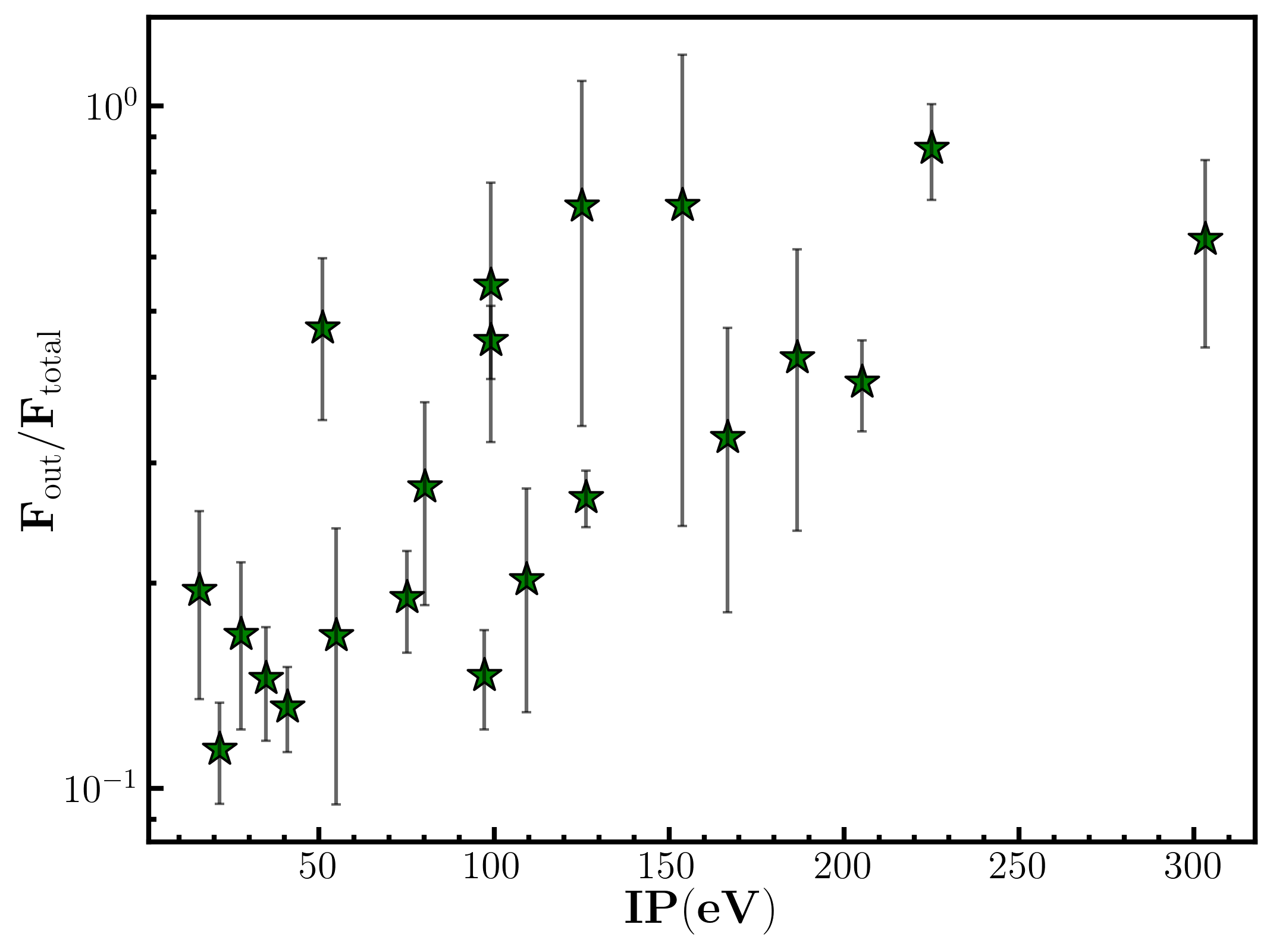}
    }
    \caption{Variation of outflow velocity ionised gas (top), outflow flux to total flux ratio (lower) (excluding [Fe II] lines) with IP.}
    \label{fig:EP-Vout-flux}
\end{figure}

We initially examined the relationship between the outflow velocity and the ionisation potential (IP) of each emission line. No statistically significant correlation was found (Pearson correlation coefficient $r = 0.20$, $p = 0.34$), suggesting that the kinematics of the outflowing gas are largely independent of the ionisation state of the emitting species. Notably, the [Fe,\textsc{ii}] lines (IP $\sim$7.9 eV) display systematically higher $V_{\mathrm{out}}$. When these [Fe,\textsc{ii}] lines are excluded, a positive correlation emerges ($r = 0.47$, $p = 0.031$) (which is shown in the upper panel of Fig. \ref{fig:EP-Vout-flux}), in agreement with previous studies \citep{2002ApJ...579..214R, 2013ApJ...777..156I}. This behaviour indicates that [Fe,\textsc{ii}], which is strongly enhanced in shocked regions, deviates from the general IP–$V_{\mathrm{out}}$ trend and likely traces a distinct shock-driven kinematic component of the outflow \citep[e.g.,][]{2001A&A...369L...5O, 2004A&A...425..457R}.

In the lower panel of Figure~\ref{fig:EP-Vout-flux}, we present the ratio of outflow flux to total line flux ($F_{\mathrm{out}}/F_{\mathrm{total}}$) as a function of IP. This ratio shows a statistically significant positive trend (Pearson correlation coefficient $r = 0.49$, $p = 0.01$) similar to the finding of \cite{2023ApJ...942L..37A} for NGC~7469, suggesting that lines from more highly ionised species tend to exhibit a stronger outflow component. This trend is becoming more prominent after excluding [Fe\,\textsc{ii}] lines with Pearson correlation coefficient $r = 0.69$, $p = 0.0006$. This trend supports the presence of an ionisation stratification in which higher-ionisation gas is more effectively coupled to outflowing motions, possibly originating from regions closer to the active nucleus where radiation pressure or mechanical feedback is stronger \citep{2018NatAs...2..198H, 2022A&A...665L..11P, 2023MNRAS.524..143F}. 
This excess in $V_{\mathrm{out}}$ and $F_{\mathrm{out}}/F_{\mathrm{total}}$ suggests a distinct excitation mechanism for [Fe\,\textsc{ii}], likely dominated by shocks rather than photoionisation, consistent with previous studies highlighting the shock sensitivity of [Fe\,\textsc{ii}] emission in AGN environments \citep[e.g.,][]{1991A&A...251...27C, 2016JKAS...49..109K, 2024ApJ...960...41M}.

\begin{figure}
    \centering
    \includegraphics[scale=0.45]{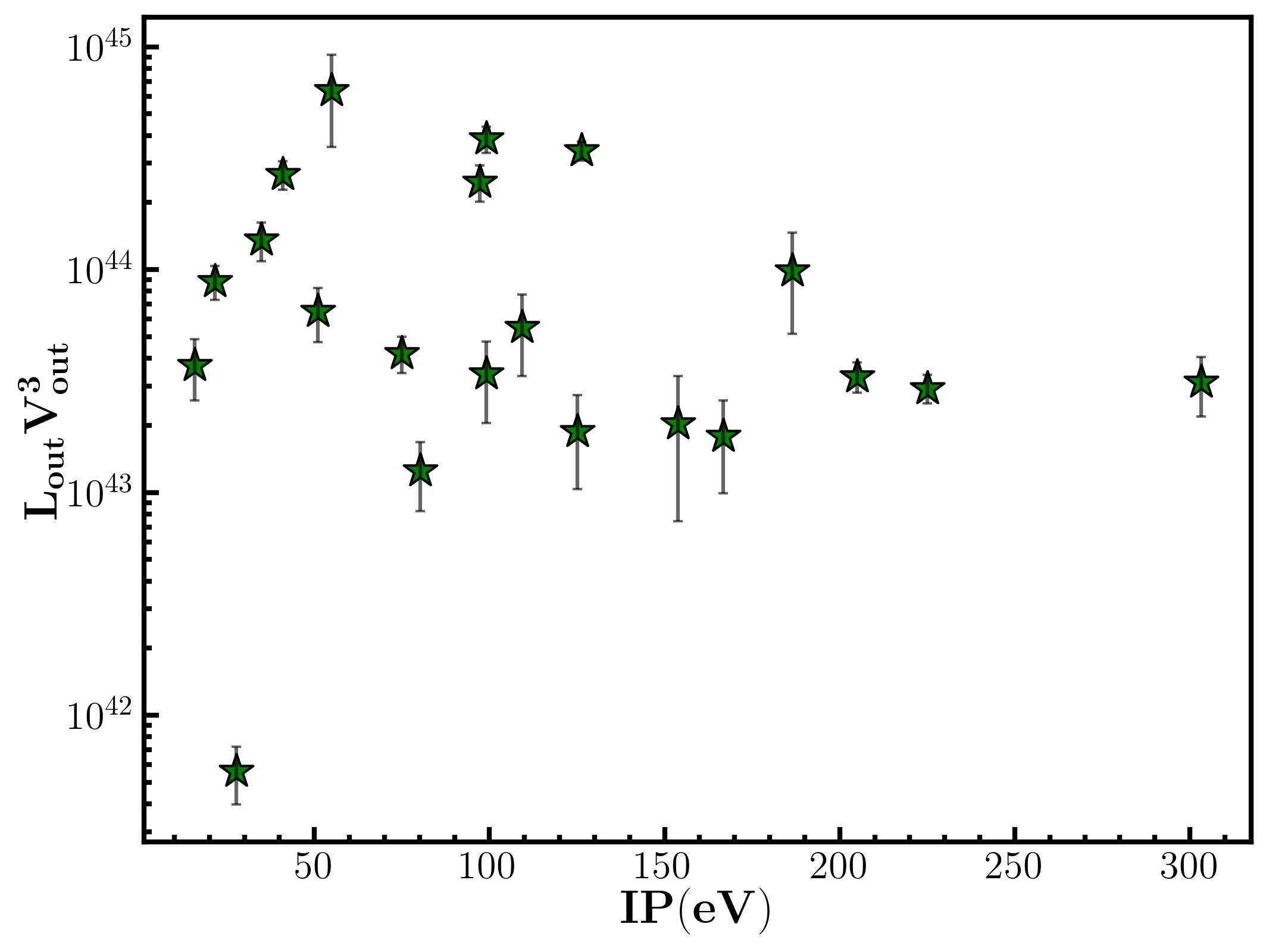}
    \caption{Variation of $L_{\mathrm{out}}V_{\mathrm{out}}^3$ with IP. $L_{\mathrm{out}}V_{\mathrm{out}}^3$ is in the unit of erg s$^{-1}$ (km s$^{-1}$)$^3$.}
    \label{fig:Lout_Vout_EP}
\end{figure}

We also attempted to assess how the outflow kinetic power might scale with ionization potential. Although a complete calculation of the kinetic power requires knowledge of the $n_e$ and $T_e$ of each ion, parameters that vary significantly with the ionization state and are not uniformly constrained for all species, an approximate indicator can be derived from $KP_{\mathrm{out}} \propto L_{\mathrm{out}} V_{\mathrm{out}}^3$. We therefore computed this proxy for each line and examined its dependence on IP. As shown in Figure~\ref{fig:Lout_Vout_EP}, the resulting trend suggests a weak negative correlation (Pearson $r = -0.260$, $p = 0.24$), with hints of decreasing $KP_{\mathrm{out}}$ at higher IP. This may reflect either a reduced mass-loading factor in more ionised gas.

Overall, these results point to a stratified, multiphase outflow where the coupling of AGN-driven forces is more effective in highly ionised gas, while the kinetic power may be distributed across different ionisation zones with varying efficiency.

The collective detection of outflows in ionized, hot molecular, warm molecular, and cold molecular gas provides compelling evidence for a multiphase outflow in NGC~4395. These components span a wide range of temperatures (from $\sim$10$^2$ to 10$^6$ K) and ionization states, underscoring the diverse impact of AGN-driven winds on the surrounding ISM. The velocity structure and line profile asymmetries vary significantly between gas phases, reflecting differences in acceleration mechanisms, gas densities, and spatial origin.
These observations demonstrate that even a low-luminosity AGN in a dwarf galaxy can drive complex, multiphase outflows, contributing to the regulation of gas dynamics and potentially influencing future star formation in the host.
\section{Summary and Conclusions}\label{sec:summary}

In this work, we present a comprehensive multi-wavelength analysis of the nearby dwarf Seyfert galaxy NGC~4395, hosting an IMBH, based on spatially resolved spectroscopy from \textit{JWST}/NIRSpec and MIRI, ALMA, and Gemini/GMOS. Our key findings are summarised as follows:

\begin{itemize}
    \item We detected 134 emission lines, including H\,\textsc{i} recombination lines (Paschen, Brackett, Pfund and Humphreys series), He\,\textsc{i} and He\,\textsc{ii} lines, fine-structure lines from both low and high-ionisation species (e.g., [Fe\,\textsc{ii}], [Ne\,\textsc{v}], [Si\,\textsc{ix}], [Mg\,\textsc{viii}]) with IP ranges from 7.6~eV to 300~eV, and pure rotational and ro-vibrational H$_2$ lines. We also detected a few distinct continuum features of PAH, 3.3 $\mu$m, 6.2 $\mu$m, 11.3 $\mu$m and a PAH plateau in the 15--20 $\mu$m region.
    
    \item Using a suite of diagnostics from [Fe\,\textsc{ii}], [Fe\,\textsc{vii}], and [Ne\,\textsc{v}], we found a clear stratification in electron density ($n_e$) and temperature ($T_e$), with higher values associated with Fe lines.

    \item We found that molecular H$_2$ gas consists of three phases of gas. The warm phase of T$\approx$580 K, the hot phase of T$\approx$1480 K and the very hot gas phase of T$\approx$2900 K.

    \item The PAH emission is dominated by large, neutral molecules, as indicated by the low PAH3.3/PAH11.3 and PAH6.2/PAH11.3 ratios and the absence of the 7.7 and 12.7\,$\mu$m bands. This pattern reflects strong AGN processing, where small PAHs are destroyed and only the more resilient neutral PAHs survive.

    \item From hydrogen recombination lines, we estimated an ionised gas mass of $M_{\rm ion} \approx$1200~M$_\odot$, under Case~B conditions and molecular hydrogen mass ($936$~M$_\odot$).
    
    \item Multi-phase outflows are detected across different ionised species through fine-structure lines, H\,\textsc{i} and molecular gas through H$_2$ lines.
    
    \item Outflow signatures appear in 36 fine-structure lines, multiple H$_2$ transitions, and the CO(2--1) line. Ionised outflows traced by both low-ionisation ([Fe\,\textsc{ii}], [Ar\,\textsc{ii}]) and high-ionisation coronal lines ([Mg\,\textsc{viii}], [Si\,\textsc{ix}]) exhibit stratified kinematics.
    Outflow velocities span 127--716~km\,s$^{-1}$, with higher velocities generally observed in high-ionisation or shock-sensitive lines such as [Fe\,\textsc{ii}], [Mg\,\textsc{viii}], and [Si\,\textsc{ix}].

    \item Cold molecular outflows are identified through kinematic decomposition of ALMA CO(2--1) emission, revealing distinct blueshifted components with lower velocity dispersions than the warm/hot molecular gas or ionised phase, but carrying a substantial gas mass component.

    \item The ionised outflow masses estimated from the H\,\textsc{i} lines are in 370--500)$~M_{\odot}$, whereas for high-ionisation tracers, we derived significantly lower outflow masses: 110--180$~M_{\odot}$.

    \item The outflow rate for cold molecular gas is 0.15--1.25 M$_\odot$ yr$^{-1}$, which is much higher than warm/hot molecular outflow (lower limit) (0.005 M$_\odot$ yr$^{-1}$) or ionised outflows (lower limit) (0.002--0.006 M$_\odot$ yr$^{-1}$ for coronal outflows and 0.01--0.03 M$_\odot$ yr$^{-1}$ for H\,\textsc{i} outflows).

   \item The kinetic coupling efficiencies for ionised outflows range from 0.4--1.4\% for the low-ionisation H\,\textsc{i} lines outflows and 0.003--0.12\% for the coronal-line outflows, well below the $\sim$1\% threshold generally considered necessary for AGN feedback to strongly suppress star formation in the host galaxy.  

    \item We found a positive correlation between both the fraction of flux in the outflowing component and outflow velocity with the IP of the emission lines, supporting the scenario of radially stratified AGN-driven feedback, where higher-ionisation gas closer to the nucleus is more efficiently accelerated. Notably, [Fe\,\textsc{ii}] lines deviate from this trend, likely reflecting their dominant excitation by shocks rather than photoionisation, consistent with their known sensitivity to shock-heated gas.

\end{itemize}
These results collectively demonstrate that even low-luminosity AGN hosted in dwarf galaxies can drive complex, multi-phase outflows and exert significant influence on their local interstellar medium. Our findings underscore the importance of high-resolution infrared and millimetre interferometry in disentangling the structure and impact of AGN feedback in low-mass galaxies.

\begin{acknowledgments}
    We thank the anonymous reviewer for their encouraging and constructive comments, which helped to improve our manuscript. This work is based [in part] on observations made with the NASA/ESA/CSA James Webb Space Telescope. The data were obtained from the Mikulski Archive for Space Telescopes at the Space Telescope Science Institute, which is operated by the Association of Universities for Research in Astronomy, Inc., under NASA contract NAS 5-03127 for JWST. These observations are associated with program ID 2016. This work is partly based on observations obtained at the Gemini Observatory, which is operated by the Association of Universities for Research in Astronomy, Inc., under a cooperative agreement with the NSF on behalf of the Gemini partnership: the National Science Foundation (United States), the Science and Technology Facilities Council (United Kingdom), the National Research Council (Canada), CONICYT (Chile), the Australian Research Council (Australia), Ministério da Ciência e Tecnologia (Brazil) and south-east CYT (Argentina). This paper makes use of the following ALMA data: ADS/JAO.ALMA\#2017.1.00572.S. ALMA is a partnership of ESO (representing its member states), NSF (USA) and NINS (Japan), together with NRC (Canada), MOST and ASIAA (Taiwan), and KASI (Republic of Korea), in cooperation with the Republic of Chile. The Joint ALMA Observatory is operated by ESO, AUI/NRAO and NAOJ. 
    This work has made use of the NASA Astrophysics Data System (ADS)\footnote{https://ui.adsabs.harvard.edu/} and the NASA/IPAC extragalactic database (NED)\footnote{https://ned.ipac.caltech.edu}. 
    PN would like to thank the COSPAR Capacity Building Programme for the support towards visiting CAB to start and complete major part of this work. 
    RAR acknowledges the support from Conselho Nacional de Desenvolvimento Cient\'ifico e Tecnol\'ogico (CNPq; Proj. 303450/2022-3, 403398/2023-1, \& 441722/2023-7) and Coordena\c c\~ao de Aperfei\c coamento de Pessoal de N\'ivel Superior (CAPES; Proj. 88887.894973/2023-00). 
    MPS acknowledges support under grants RYC2021-033094-I, CNS2023-145506, and PID2023-146667NB-I00 funded by MCIN/AEI/10.13039/501100011033 and the European Union NextGenerationEU/PRTR. J.A.-M. acknowledges support by grants PIB2021-127718NB-100 \& PID2024-158856NA-I00 from the Spanish Ministry of Science and Innovation/State Agency of Research MCIN/AEI/10.13039/501100011033 and by “ERDF A way of making Europe".
\end{acknowledgments}

\begin{contribution}
The first author conducted the data reduction and analysis and led the writing of the manuscript. The co-authors contributed to the interpretation of the results and provided detailed, constructive feedback that shaped the manuscript into its current form.

\end{contribution}

\vspace{5mm}
\facilities{JWST (NIRSpec and MIRI), ALMA, GEMINI (GMOS)}

\software{Numpy \citep{harris2020array}, 
Astropy \citep{2022ApJ...935..167A, 2013A&A...558A..33A, 2018AJ....156..123A}, 
Scipy \citep{2020SciPy-NMeth}, Matplotlib \citep{Hunter:2007}, PyNeb \citep{2015A&A...573A..42L}, PDR \cite{2023AJ....165...25P, 2008ASPC..394..654P, 2006ApJ...644..283K, 2011ascl.soft02022P}         }

\pagebreak
\appendix


\restartappendixnumbering 
\section{Instrumental resolution} \label{appen:resolution}
The instrumental resolutions of NIRSpec and MIRI are compared with the measured linewidths, shown as labelled data points in the figure. In the left panel, the dotted curve represents our measurements, while the solid curves correspond to previously reported MIRI spectral resolutions from \citet{2023A&A...675A.111A} and \citet{2024ApJ...963..158P, 2025AJ....169..165B}.  
In the right panel, we compare the resolving power derived from our observed NIRSpec lines (shown as dotted data points) with the expected instrumental resolution indicated by the blue and red curves, along with their shaded regions, which represent the 30\% uncertainty reported on the JWST--NIRSpec documentation page


\begin{figure}[h]
    \centering
    \hbox{
    \includegraphics[scale=0.4]{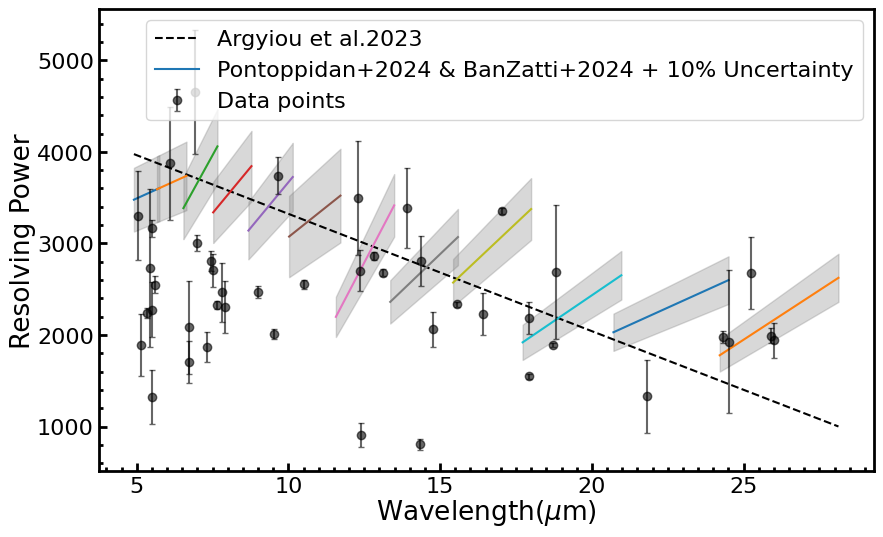}
    \includegraphics[scale=0.4]{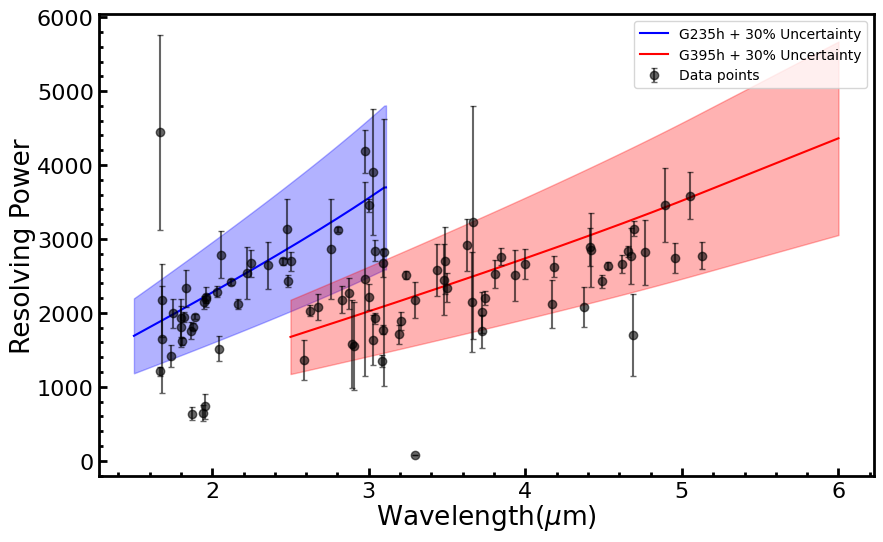}
    }
    \caption{Variation of spectral resolving power with wavelength in MIRI (left panel) and NIRSpec (right panel).}
    \label{fig:res}
\end{figure}

\section{Diagnosis for electron density and temperature} \label{appen:density-temp}
Here we examine a range of emission line diagnostics to evaluate their reliability as tracers of n$_e$ and T$_e$, identifying which indicators provide robust constraints and which are less effective under the physical conditions probed in this study.

\begin{figure}[ht]
    \centering
    \hbox{
    \includegraphics[scale=0.3]{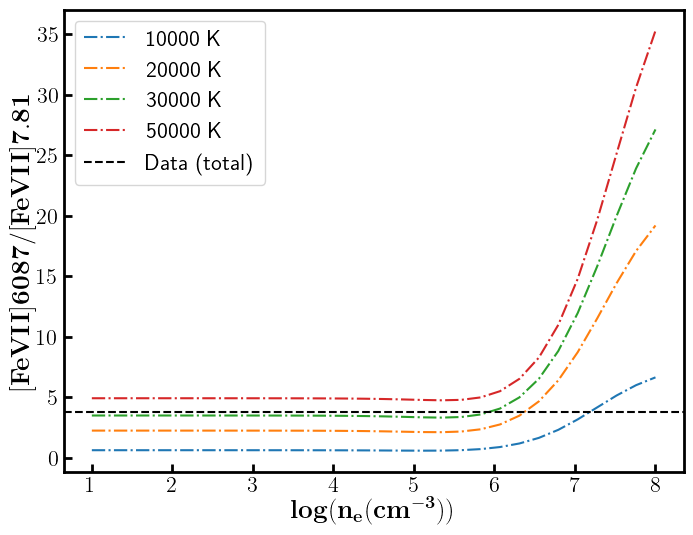}
    \includegraphics[scale=0.3]{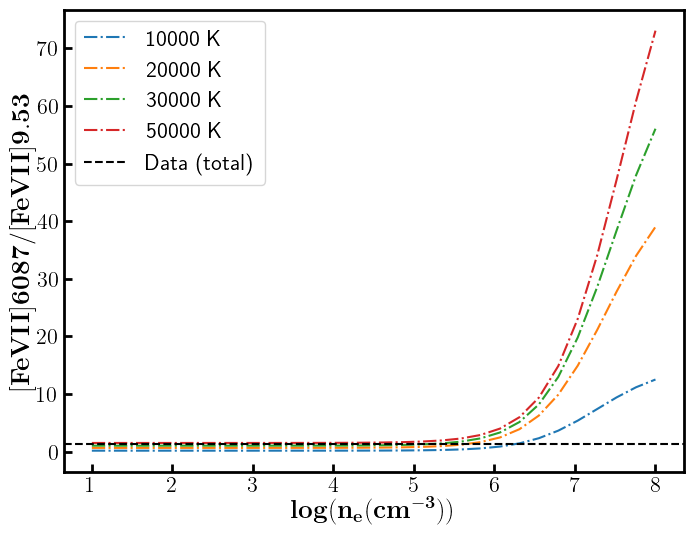}
    \includegraphics[scale=0.3]{images/ne_Te/Fe_VII_7.81_9.53_emiss_pyneb.png}
    }
    \caption{Variation of different line ratios with density for different particular temperatures, which represent different curves in each plot and are denoted by individual legends in each plot. The horizontal line in each plot represents the observed line ratio.}
    \label{fig:Appen-density}
\end{figure}

\begin{figure}[ht]
    \hbox{
    \includegraphics[scale=0.3]{images/ne_Te/Fe_VII_6087_7.81_emiss_tempera_pyneb.png}
    \includegraphics[scale=0.3]{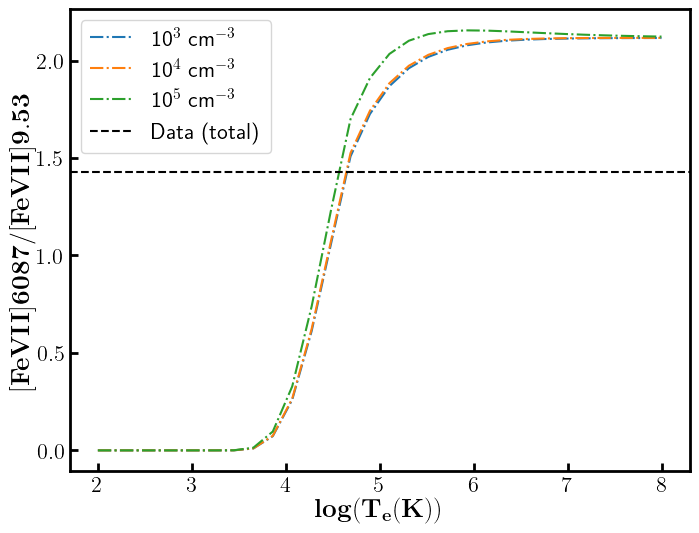}
    \includegraphics[scale=0.3]{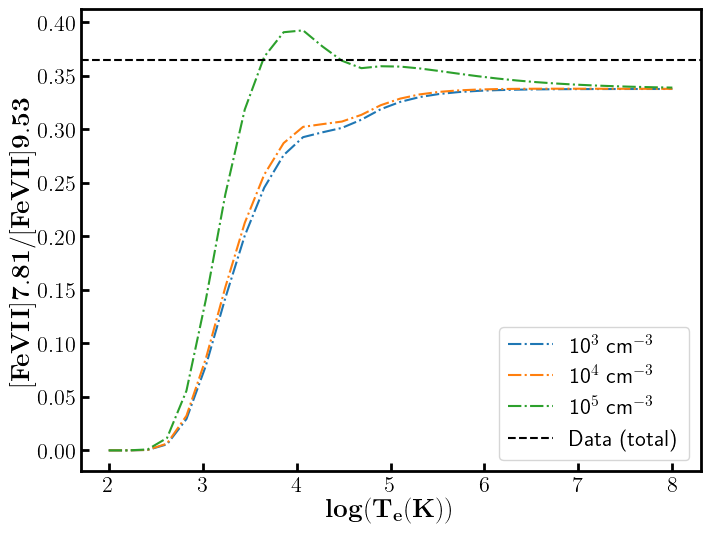}
    }
    \caption{Variation of different line ratios with electron temperature for different particular electron densities, which represent different curves in each plot and are denoted by individual legends in each plot. The horizontal line in each plot represents the observed line ratio.}
    \label{fig:Appen-temp}
\end{figure}

\newpage
\section{Identified lines}
Here, we listed all the identified emission lines together with their IP, wavelengths, and flux in the nuclear spectra.
\begin{longtable}{lcccc}
\caption{Emission line measurements identified in the nuclear spectra.The Flux and Flux error are in unit of $10^{-16}$ erg s$^{-1}$ cm$^{-2}$.\label{tab:line-list}}\\
\hline\hline
Line & IP (eV) & $\lambda_{\rm vac}$ ($\mu$m) & Flux $_{\rm total}$ & Flux $_{\rm total}^{err}$  \\
\hline
\endfirsthead
\caption{continued.}\\
\hline\hline
Line & IP (eV) & $\lambda_{\rm vac}$ ($\mu$m) & Flux $_{\rm total}$  & Flux $_{\rm total}^{err}$ \\
\hline
\endhead
\hline
\endfoot
\hline
\endlastfoot

\multicolumn{4}{c}{\textbf{H\,\textsc{I} lines}} \\
H\,I\,(11--4) & -- & 1.6811 & 4.2 & 1.6 \\
H\,I\,(10--4) & -- & 1.7367 & 6.7 & 1.3 \\
H\,I\,(9--4)  & -- & 1.8179 & 7.9 & 0.5 \\
H\,I\,(4--3)  & -- & 1.8756 & 213.2 & 8.0 \\
H\,I\,(8--4)  & -- & 1.9451 & 11.4 & 0.9 \\
H\,I\,(7--4)  & -- & 2.1661 & 17.6 & 0.9 \\
H\,I\,(6--4)  & -- & 2.6259 & 28.1 & 1.8 \\
H\,I\,(13--5) & -- & 2.6744 & 1.0 & 0.1 \\
H\,I\,(12--5) & -- & 2.7583 & 2.1 & 0.7 \\
H\,I\,(11--5) & -- & 2.8730 & 2.9 & 0.5 \\
H\,I\,(10--5) & -- & 3.0392 & 3.7 & 0.4 \\
H\,I\,(9--5)  & -- & 3.2970 & 4.3 & 1.0 \\
H\,I\,(8--5)  & -- & 3.7406 & 6.6 & 0.6 \\
H\,I\,(13--6) & -- & 4.1708 & 0.7 & 0.1 \\
H\,I\,(12--6) & -- & 4.3765 & 1.2 & 0.3 \\
H\,I\,(7--5)  & -- & 4.6538 & 9.1 & 0.4 \\
H\,I\,(11--6) & -- & 4.6725 & 2.0 & 0.4 \\
H\,I\,(10--6) & -- & 5.1287 & 1.9 & 0.5 \\
H\,I\,(9--6)  & -- & 5.9082 & 3.2 & 0.9 \\
H\,I\,(6--5)  & -- & 7.4599 & 15.2 & 1.2 \\
H\,I\,(8--6)  & -- & 7.5025 & 4.5 & 0.6 \\
H\,I\,(11--7) & -- & 5.5081 & 1.1 & 0.3 \\
H\,I\,(7--6)  & -- & 12.372 & 6.7 & 1.0 \\
H\,I\,(11--8) & -- & 12.387 & 2.4 & 0.4 \\
\multicolumn{4}{c}{\textbf{He lines}} \\
He\,II       & -- & 1.8642 &  8.9 & 0.8 \\
He\,I        & -- & 1.8690 & 12.0 & 2.2 \\
He\,I        & -- & 2.0587 & 2.5 & 0.4 \\
He\,II       & -- & 2.8260 & 1.2 & 0.9 \\
He\,II       & -- & 3.0908 & 4.3 & 0.5 \\
He\,II       & -- & 3.0917 & 3.9 & 0.2 \\
He\,II       & -- & 3.0955 & 0.2 & 0.1 \\
He\,II       & -- & 4.7635 & 1.8 & 0.8 \\
\multicolumn{4}{c}{\textbf{H$_2$ lines}} \\
H$_2$(1--0) S(7)  & -- & 1.7481 & 2.3 & 0.1 \\
H$_2$(1--0) S(6)  & -- & 1.7880 & 1.6 & 0.1 \\
H$_2$(1--0) S(5)  & -- & 1.8358 & 4.3 & 1.1 \\
H$_2$(1--0) S(4)  & -- & 1.8920 & 1.79 & 0.04 \\
H$_2$(1--0) S(3)  & -- & 1.9576 & 7.4 & 0.3 \\
H$_2$(1--0) S(2)  & -- & 2.0338 & 3.0 & 0.1 \\
H$_2$(1--0) S(1) & -- & 2.1218 & 8.8 & 0.2 \\
H$_2$(1--0) S(0) & -- & 2.2235 & 3.3 & 0.7 \\
H$_2$(2--1) S(1) & -- & 2.2477 & 1.2 & 0.1 \\
H$_2$(2--1) S(0) & -- & 2.3556 & 0.6 & 0.1 \\
H$_2$(1--0) Q(5) & -- & 2.4549 & 4.0 & 0.1 \\
H$_2$(1--0) Q(6) & -- & 2.4756 & 1.1 & 0.2 \\
H$_2$(1--0) Q(7) & -- & 2.5001 & 1.6 & 0.1 \\
H$_2$(1--0) O(3) & -- & 2.8025 & 7.8 & 0.2 \\
H$_2$(2--1) O(3) & -- & 2.9741 & 1.0 & 0.3 \\
H$_2$(1--0) O(4) & -- & 3.0038 & 2.3 & 0.8 \\
H$_2$(3--2) O(3) & -- & 3.1637 & 0.5 & 0.1 \\
H$_2$(2--1) O(4) & -- & 3.1899 & 0.7 & 0.1 \\
H$_2$(1--0) O(5) & -- & 3.2349 & 3.7 & 0.1 \\
H$_2$(2--1) O(5) & -- & 3.4379 & 0.7 & 0.2 \\
H$_2$(1--0) O(6) & -- & 3.5007 & 1.0 & 0.1 \\
H$_2$(0--0) S(15) & -- & 3.6264 & 0.7 & 0.2 \\
H$_2$(3--2) O(5) & -- & 3.6632 & 0.16 & 0.03 \\
H$_2$(2--1) O(6) & -- & 3.7237 & 0.6 & 0.1 \\
H$_2$(0--0) S(14) & -- & 3.7256 & 0.5 & 0.1 \\
H$_2$(0--0) S(13) & -- & 3.8472 & 1.1 & 0.1 \\
H$_2$(0--0) S(12) & -- & 3.9969 & 0.4 & 0.03 \\
H$_2$(1--0) O(7) & -- & 3.8074 & 1.2 & 0.1 \\
H$_2$(0--0) S(11) & -- & 4.1815 & 1.9 & 0.1 \\
H$_2$(0--0) S(10) & -- & 4.4099 & 1.0 & 0.2 \\
H$_2$(1--1) S(11) & -- & 4.4167 & 0.4 & 0.1 \\
H$_2$(0--0) S(9) & -- & 4.6947 & 4.1 & 0.3 \\
H$_2$(1--1) S(9) & -- & 4.9542 & 0.5 & 0.1 \\
H$_2$(0--0) S(8) & -- & 5.0531 & 2.2 & 0.6 \\
H$_2$(0--0) S(7) & -- & 5.5118 & 11.3 & 0.5 \\
H$_2$(0--0) S(6) & -- & 6.1086 & 7.4 & 0.4 \\
H$_2$(0--0) S(5)  & -- & 6.9095 & 42.8 & 9.7 \\
H$_2$(0--0) S(4) & -- & 8.0251 & 25.7 & 9.3 \\
H$_2$(0--0) S(3) & -- & 9.6649 & 110.2 & 9.1 \\
H$_2$(0--0) S(2) & -- & 12.279 & 59.6 & 5.3 \\
H$_2$(0--0) S(1) & -- & 17.035 & 100.9 & 2.3 \\
\multicolumn{4}{c}{\textbf{Fine structure lines}} \\
\,[Fe II\,]       & 7.9    & 1.6642 & 1.8 & 0.1 \\
Fe I\,]           & --     & 1.6684 & 2.2 & 0.7 \\
\,[Fe II\,]       & 7.9    & 1.6773 & 3.9 & 0.6 \\
\,[Fe II\,]       & 7.9    & 1.7976 & 1.3 & 0.3 \\
\,[Fe II\,]       & 7.9    & 1.8005 & 2.1 & 0.7 \\
\,[Fe II\,]       & 7.9    & 1.8094 & 7.3 & 0.5 \\
\,[Ni II\,]       & 7.64   & 1.9393 & 2.4 & 0.5 \\
\,[Fe II\,]       & 7.9    & 1.9541 & 1.2 & 0.4 \\
\,[Si VI\,]       & 166.7  & 1.9634 & 11.5 & 1.8 \\
\,[Fe II\,]       & 7.9    & 2.0465 & 1.0 & 0.2 \\
\,[Si VII\,]      & 205.05 & 2.4833 & 9.6 & 0.6 \\
\,[Si IX\,]       & 303.17 & 2.5842 & 1.7 & 0.2 \\
\,[Mg VIII\,]     & 224.95 & 3.0279 & 3.1 & 0.3 \\
\,[Fe V\,]        & 54.8   & 2.8934 & 0.8 & 0.3 \\
\,[Fe III\,]      & 16.18  & 2.9049 & 0.7 & 0.3 \\
\,[Mg VIII\,]     & 224.95 & 3.0279 & 2.2 & 0.8 \\
\,[Fe II\,]       & 7.9    & 3.0816 & 1.9 & 0.1 \\
\,[Ca IV\,]       & 50.91  & 3.2067 & 5.0 & 0.6 \\
Fe I\,]           & --     & 3.4831 & 0.27 & 0.07 \\
Mg I\,]           & --     & 3.4856 & 0.33 & 0.07 \\
\,[Al VI\,]       & 153.83 & 3.6597 & 1.7 & 0.7 \\
\,[Si IX\,]       & 303.17 & 3.9357 & 1.9 & 0.3 \\
\,[Mg IV\,]       & 80.14  & 4.4867 & 8.1 & 0.7 \\
\,[Ar VI\,]       & 75.02  & 4.5295 & 28.9 & 1.1 \\
\,[K III\,]       & 31.63  & 4.6180 & 1.5 & 0.1 \\
\,[Na VII\,]      & 172.15 & 4.6847 & 0.23 & 0.1 \\
\,[Fe II\,]       & 7.9    & 4.8891 & 0.70 & 0.2 \\
\,[Fe II\,]       & 7.9    & 5.3402 & 28.6 & 1.7 \\
\,[Fe VIII\,]     & 125.1  & 5.4466 & 5.6 & 1.8 \\
\,[Mg VII\,]      & 186.51 & 5.5032 & 5.8 & 1.2 \\
\,[Mg V\,]        & 109.24 & 5.6098 & 20.5 & 1.7 \\
\,[Ni II\,]       & 7.64   & 6.636  & 3.8 & 0.2 \\
\,[Cl V\,]        & 53.46  & 6.7067 & 0.8 & 0.3 \\
\,[Fe II\,]       & 7.9    & 6.7213 & 1.6 & 0.3 \\
\,[Ar II\,]       & 15.76  & 6.9853 & 198.9 & 14.3 \\
\,[Na III\,]      & 47.29  & 7.3177 & 1.6 & 0.2 \\
\,[Ne VI\,]       & 126.21 & 7.6524 & 105.3 & 3.4 \\
\,[Fe VII\,]      & 99.1   & 7.8145 & 4.6 & 0.3 \\
\,[Ar V\,]        & 59.81  & 7.9016 & 11.7 & 1.5 \\
\,[Ar III\,]      & 27.63  & 8.9914 & 0.68 & 0.04 \\
\,[Fe VII\,]      & 99.1   & 9.5267 & 12.5 & 0.4 \\
\,[S IV\,]        & 34.79  & 10.511 & 239.5 & 9.1 \\
\,[Ne II\,]       & 21.56  & 12.814 & 550.3 & 16.8 \\
\,[Ar V\,]        & 59.81  & 13.102 & 16.8 & 2.7 \\
\,[Mg V\,]        & 109.24 & 13.521 & 1.6 & 0.6 \\
\,[Ne V\,]        & 97.12  & 14.322 & 199.7 & 6.7 \\
\,[Fe VI\,]       & 75.0   & 13.906 & 0.58 & 0.15 \\
\,[Cl II\,]       & 12.97  & 14.368 & 4.9 & 0.6 \\
\,[Fe VI\,]       & 75.0   & 14.771 & 2.2 & 0.3 \\
\,[Ne III\,]      & 40.96  & 15.555 & 764.2 & 19.3 \\
\,[Co III\,]      & 17.08  & 16.407 & 0.94 & 0.28 \\
\,[Fe II\,]       & 7.9    & 17.936 & 22.9 & 0.6 \\
\,[S III\,]       & 23.34  & 18.713 & 308.5 & 8.5 \\
\,[Co II\,]       & 17.08  & 18.804 & 0.97 & 0.33 \\
\,[Ar III\,]      & 27.63  & 21.830 & 4.5 & 0.8 \\
\,[Fe III\,]      & 16.19  & 22.925 & 4.7 & 0.7 \\
\,[Ne V\,]        & 97.12  & 24.318 & 156.0 & 9.8 \\
\,[Fe II\,]       & 7.9    & 24.519 & 7.7 & 0.9 \\
\,[S I\,]         & --     & 25.249 & 5.9 & 1.2 \\
\,[O IV\,]        & 54.93  & 25.890 & 728.9 & 65.7 \\
\,[Fe II\,]       & 7.9    & 25.988 & 31.0 & 4.3 \\
\end{longtable}
\tablecomments{
IP = Ionization potential of the corresponding ion; $\lambda_{\rm vac}$ = vacuum wavelength of the transition.}

\clearpage

\bibliography{ref}{}
\bibliographystyle{aasjournalv7}

\end{document}